\documentclass[12pt]{article}
\pdfoutput=1
\usepackage[nosort]{cite}

\usepackage{xcolor}

\usepackage{epsfig}
\usepackage{amsfonts}
\usepackage{amscd}
\usepackage{latexsym}
\usepackage{amsmath,amssymb}
\usepackage{verbatim}
\usepackage{setspace}
\usepackage{color}
\usepackage{fancyhdr}
\usepackage{cite}
\usepackage{hyperref}
\usepackage{tikz}
\usepackage{slashed}
\usetikzlibrary{calc}
\usetikzlibrary{topaths}
\usetikzlibrary{decorations}
\usetikzlibrary{decorations.pathmorphing}

\usepackage{draft}
\usepackage{hyperref}
\usepackage{graphicx,color,subfig}
\usepackage{cite}
\usepackage{mciteplus}
\usepackage{skak}

\numberwithin{equation}{section}

\def\d{\partial}

\begin{document}
	
\begin{titlepage}
	
	\begin{center}
		
		\hfill \\
		\hfill \\
		\vskip 1cm

		\title{The low-energy limit of some exotic lattice theories and UV/IR mixing}
		
		\author{Pranay Gorantla$^{1}$, Ho Tat Lam$^{1}$, Nathan Seiberg$^{2}$ and Shu-Heng Shao$^{2}$}
		
		\address{${}^{1}$Physics Department, Princeton University, Princeton NJ, USA}
		\address{${}^{2}$School of Natural Sciences, Institute for Advanced Study, Princeton NJ, USA}

	\end{center}
	
	\vspace{2.0cm}
	
	\begin{abstract}\noindent
		We continue our exploration of exotic, gapless lattice and continuum field theories with subsystem  global symmetries.
		In an earlier paper, we presented free lattice models enjoying all the global symmetries (except continuous translations), dualities, and anomalies of the  continuum theories.
		Here, we study in detail the relation between the lattice models and the corresponding continuum theories.  We do that by analyzing the spectrum of the theories and  several correlation functions.  These lead us to uncover interesting subtleties in the way the continuum limit can be taken.  In particular, in some cases, the infinite volume limit and the continuum limit do not commute.  This signals a surprising UV/IR mixing, i.e., long distance sensitivity to short distance details.
	\end{abstract}
	
	\vfill
	
\end{titlepage}

\eject

\tableofcontents

\section{Introduction}

Over the past several years, various exotic lattice models have been discovered and analyzed.  These include the gapless theory of \cite{PhysRevB.66.054526}, the gapped fracton models of \cite{PhysRevLett.94.040402,PhysRevA.83.042330}, and a lot of followup papers.

In general, a useful guiding principle in organizing and analyzing subtle systems is to focus on their global symmetries and their consequences.  Our discussion here follows our earlier papers \cite{Seiberg:2019vrp,paper1,paper2,paper3,Gorantla:2020xap,Gorantla:2020jpy,Rudelius:2020kta,Gorantla:2021svj} where the global symmetries of these exotic systems were discussed in detail.  This led to the realization that the key features in these models is the existence of a subsystem global symmetry.\footnote{Subsystem global symmetries have appeared in various earlier papers including e.g., \cite{PhysRevLett.85.2160}.}
This is a huge global symmetry, whose size grows with the number of sites in the lattice.  This global symmetry could be present in the UV lattice model, or it could emerge in the IR, as an accidental symmetry.
The existence of such a large global symmetry has several interesting consequences, including the following.
\begin{itemize}
\item The same operator at different locations can carry different charges under this symmetry.  Consequently,  correlation functions of charged operators change rapidly as their positions change at the lattice scale.  In the continuum limit, these correlation functions become discontinuous.
\item The large global symmetry can have large representations.  In particular, the ground state of the system can be in such a large representation.  This leads to a large ground state degeneracy, which can diverge in the continuum limit.\footnote{In fact, depending on the details of the global symmetry, the ground state degeneracy might not even be monotonic in the system size \cite{PhysRevA.83.042330,Shirley:2017suz,Slagle:2018wyl,Dua:2019ucj,1821601,Meng,Manoj:2020wwy,Rudelius:2020kta}.}
\end{itemize}

These facts, which follow from the underlying subsystem global symmetry, make the long-distance behavior of these systems sensitive to some short-distance phenomena. This is reminiscent of the {\it UV/IR mixing} in quantum field theory on a noncommutative space \cite{Minwalla:1999px}.  This UV/IR mixing is at the roots of the difficulties in describing these systems using conventional continuum quantum field theory \cite{paper1,paper2,paper3,Gorantla:2020xap,Slagle:2020ugk,Gorantla:2020jpy,Rudelius:2020kta, You:2021tmm,Gorantla:2021svj,Zhou:2021wsv,Hsin:2021mjn,You:2021sou,Casalbuoni:2021fel}.

In order to avoid confusion, we emphasize that the UV/IR mixing is not a mixing between low energies and high energies.  Instead, it is mixing between small momenta and low energies on one side and high momenta on the other side.

This UV/IR mixing is also manifest in other aspects of the theories.    As we will see, in the continuum limit, some correlation functions diverge at non-coincident points.  Such correlation functions are finite once they are regulated in the UV by a nonzero lattice spacing $a$.  Alternatively, they are finite once they are regulated in the IR by a finite physical length $\ell$.

We will start by focusing on the simplest model with such a symmetry: the 2+1d XY-plaquette model of \cite{PhysRevB.66.054526}.  The degrees of freedom are circle-valued $\phi$ at the sites of a two-dimensional lattice and the Hamiltonian is
\ie\label{eq:iH_XY_plaq}
H  = { \mathcal{U} \over2} \sum_{\text{site}} \pi^2 - {\mathcal{K} }\sum_{xy-\text{plaq}}\cos(\Delta_x\Delta_y\phi)~,
\fe
where $\pi$ is the momentum conjugate to $\phi$.   This theory enjoys a subsystem global symmetry \cite{PhysRevB.66.054526}, which, following the standard string theory terminology, we refer to as ``momentum subsystem symmetry'' \cite{paper1}.  In the condensed matter literature, this symmetry is referred to as ``particle number symmetry.''

For ${\mathcal{K}/ \mathcal{U}}$ large enough,  the XY-plaquette model is in a gapless phase, which can be described by the Euclidean continuum action \cite{PhysRevB.66.054526}
\ie\label{eq:i action_continuum}
S=\int d\tau dx dy\left[\frac{\mu_0}{2}(\partial_\tau\phi)^2+\frac{1}{2\mu}(\partial_x\partial_y\phi)^2\right]~.
\fe

The relation between the parameters in \eqref{eq:iH_XY_plaq} and in \eqref{eq:i action_continuum} is obtained as follows.
The Euclidean action (with continuous Euclidean time) corresponding to the Hamiltonian  \eqref{eq:iH_XY_plaq} is
\ie\label{eq:iS_XY_plaq}
{ 1 \over2 \mathcal{U}} \int d\tau \sum_{\text{site}} (\partial_\tau \phi)^2 - {\mathcal{K} }\int d\tau \sum_{xy-\text{plaq}}\cos(\Delta_x\Delta_y\phi)~.
\fe
The continuum limit is obtained for $\mathcal K\gg \mathcal U$. (Below we will analyze it in more detail.)  Then, we can assume that $\Delta_x\Delta_y \phi \ll 1$ and replace $\cos(\Delta_x\Delta_y\phi)\approx 1-{1\over 2} (\Delta_x\Delta_y\phi)^2$.
Next, we introduce a spatial lattice spacing $a$ and write \eqref{eq:iS_XY_plaq} as
\ie\label{eq:iS_XY_plaq c}
 { 1 \over2 \mathcal Ua^2} \int d\tau dx dy~ (\partial_\tau \phi)^2 + {\mathcal K a^2\over 2}  \int d\tau dx dy~  (\partial_x\partial_y\phi)^2~,
\fe
where we dropped an additive constant.  Therefore, we identify the continuum limit as the limit $a\to 0$ with $\mathcal U,\mathcal K\to \infty$, such that
\ie\label{mathcalUKmumuz}
&\mu_0={1\over \mathcal Ua^2}~, \\
&\mu={1\over \mathcal Ka^2}~,
\fe
are held fixed.  Since we assumed above that $\mathcal K\gg \mathcal U$, this is valid only for $\mu\ll \mu_0$.  Yet, we will be interested in the theory based on \eqref{eq:i action_continuum} for all values of $\mu $ and $\mu_0$. Soon, we will discuss another lattice model that leads to \eqref{eq:i action_continuum} with arbitrary $\mu$ and $\mu_0$.

The theory based on the continuum action \eqref{eq:i action_continuum} has been analyzed in  \cite{PhysRevB.66.054526,Tay_2011,You:2019cvs,You:2019bvu,paper1,Karch:2020yuy,Rudelius:2020kta,You:2021tmm,Casalbuoni:2021fel}.  In particular, \cite{paper1} analyzed in detail many of the properties of this continuum theory.  These include exotic global symmetries, a peculiar spectrum, subtle correlation functions, and a surprising self-duality.
  Specifically, in addition to the momentum subsystem symmetry of the lattice theory \eqref{eq:iH_XY_plaq} \eqref{eq:iS_XY_plaq}, the continuum theory \eqref{eq:i action_continuum} also has another subsystem symmetry.  Again, following the string theory terminology, we refer to it as a ``winding subsystem symmetry'' \cite{paper1}.

The continuum action \eqref{eq:i action_continuum} also has two independent emergent scaling symmetries \cite{Gromov:2018nbv,Karch:2020yuy}
\ie\label{scalingxy}
&x\to \lambda_x x ~,\\
&y\to \lambda_y y ~,\\
&\tau\to \lambda_x \lambda_y \tau ~.\\
\fe
These scaling symmetries are not present in any of the related lattice models.  When we formulate the system on a torus with lengths $\ell_x$ and $\ell_y$ in the $x$ and $y$ directions, this scaling symmetry acts on these lengths in an obvious way.  Finally, we note that $\mu $ and $\mu_0$ are dimensionful.  Therefore, the continuum physics depends only on the three dimensionless parameters $\mu/\mu_0$, $\ell_x/\ell_y$ and $\mu\ell_x$.

In this paper, we will clarify the relation between the lattice system \eqref{eq:iH_XY_plaq} and its continuum limit \eqref{eq:i action_continuum}.  In addition, since the analysis of the continuum theory in \cite{paper1} was quite subtle, and some of its conclusions might appear questionable, we will repeat it here from another perspective, placing it on significantly firmer footing.

The key point of our analysis is to replace the lattice Hamiltonian model \eqref{eq:iH_XY_plaq}, or its Euclidean action \eqref{eq:iS_XY_plaq}, by the modified Villain Euclidean action \cite{Gorantla:2021svj}\footnote{Modified Villain actions of ordinary lattice models were motivated by various papers, and in particular \cite{Gross:1990ub}.  They were discussed in detail in \cite{Sulejmanpasic:2019ytl,Gorantla:2021svj}. In our case, the Euclidean action \eqref{eq:iS_XY_plaq}, can be written in Villain form.  It is given by the first two terms in \eqref{eq:iS-Villain-lattice}.  The third term in \eqref{eq:iS-Villain-lattice} has the effect of ``suppressing the vortices.'' }
\ie\label{eq:iS-Villain-lattice}
S=\frac{\beta_0}{2}\sum_{\tau\text{-link}}(\Delta_\tau\phi-2\pi n_\tau)^2+\frac{\beta}{2}\sum_{xy\text{-plaq}}(\Delta_x\Delta_y\phi-2\pi n_{xy})^2+i\sum_{\text{cube}}\phi^{xy}(\Delta_\tau n_{xy}-\Delta_x\Delta_y n_\tau)~.
\fe
Here, $\phi$ is an $\mathbb{R}$-valued field on the sites of the Euclidean lattice and $(n_\tau,n_{xy})$ are $\mathbb{Z}$-valued tensor gauge fields on $\tau$-links and $xy$-plaquettes.  The gauge symmetry is
\ie
&\phi\sim\phi + 2\pi k~,
\\
&(n_\tau,n_{xy})\sim (n_\tau+\Delta_\tau k,n_{xy}+\Delta_x\Delta_y k)~,
\fe
where $k$ is an integer. $\phi^{xy}$ is a circle-valued Lagrange multiplier field on the cubes, which constrains the tensor gauge field $(n_\tau,n_{xy})$ to be flat.\footnote{The field $\phi^{xy}$ is closely related to the field $\theta$ in \cite{PhysRevB.66.054526}.}

Since the action \eqref{eq:iS-Villain-lattice} is quadratic, we can easily find effective actions that lead to the same correlation functions for widely separated operators.  For correlation functions of operators separated by many sites in the time direction, we can replace the Euclidean lattice with a spatial lattice and a continuous Euclidean time.  We do that by introducing a lattice spacing in the time direction $a_\tau$ and replacing the integer time coordinate by a continuous one.  Ignoring the integer-valued gauge fields, we find the effective action
\ie\label{eq:iS-Villain-continuum-time}
S&=\int d\tau\left[ \frac{1}{2U}\sum_{\text{site}}(\partial_\tau\phi)^2+\frac{K}{2}\sum_{xy-\text{plaq}}(\Delta_x\Delta_y\phi)^2\right]~,
\fe
where the parameters are
\ie
U=\frac{1}{\beta_0 a_\tau},\quad K =\frac{\beta}{a_\tau}~.
\fe
We emphasize that this effective action is valid for all values of $\beta$ and $\beta_0$, provided the operators are widely separated in the time direction.

We can repeat this process in the spatial directions. We introduce a lattice spacing $a$ and for correlation functions of operators that are widely separated in the space directions, we can consider the effective continuum action
\ie
\frac{1}{2Ua^2}\int d\tau dxdy (\partial_\tau\phi)^2+\frac{K a^2}{2}\int d\tau dxdy (\partial_x\partial_y\phi)^2~.
\fe
(Below we will discuss it in more detail.)  Now, we identify
\ie\label{UKmumuzbeta}
&\mu_0={1\over Ua^2}= {\beta_0 a_\tau \over a^2}~, \\
&\mu={1\over Ka^2}={a_\tau \over \beta a^2}~.
\fe
Hence, the continuum limit corresponds to $a,a_\tau\to 0$ with fixed $\mu,\mu_0$.  The scaling symmetries \eqref{scalingxy} motivate us to consider the continuum limit with $a_\tau/ a^2$ held fixed.  Then, $\beta$ and $\beta_0$ are held fixed in the continuum limit.  Again, we will discuss it in more detail below.

This paper is the logical continuation of \cite{Gorantla:2021svj}.  In \cite{Gorantla:2021svj}, we presented the model \eqref{eq:iS-Villain-lattice} and analyzed its kinematical features.  Here, we explore its dynamical aspects.\footnote{In high energy physics, it is common to refer to the symmetries, their anomalies, and their consequences as ``kinematics'', and to the effects specific to the Hamiltonian or the Lagrangian as ``dynamics.''  Here we follow this terminology.  We do not mean ``dynamics'' in the sense of time dependence.}

The Euclidean lattice model \eqref{eq:iS-Villain-lattice} provides a nice bridge between the lattice Hamiltonian system \eqref{eq:iH_XY_plaq} and the continuum action \eqref{eq:i action_continuum}. One aspect of this bridge is that the modified Villain theory \cite{Gorantla:2021svj} has both the momentum and the winding subsystem symmetries of the continuum theory. (Of course, it lacks the continuous spacetime translation symmetry and the scaling symmetries \eqref{scalingxy} of the continuum theory.)

As discussed in \cite{Gorantla:2021svj}, the Euclidean lattice model \eqref{eq:iS-Villain-lattice} leads to the same physics as the lattice Hamiltonian \eqref{eq:iH_XY_plaq}, provided $\mathcal K\gg\mathcal U$.  This will be reviewed in Section \ref{sec:2+1d-XYplaq}.  Note the similarities between \eqref{mathcalUKmumuz} and \eqref{UKmumuzbeta}.  For $K\gg U$, we have $K\approx {\mathcal K}$ and $U\approx {\mathcal U}$ and all our different theories coincide.  In this paper, we will be interested in generic values of $\beta$ and $\beta_0$ and correspondingly generic $K/ U$.  Then \eqref{mathcalUKmumuz} might not be valid, but \eqref{UKmumuzbeta} is still correct.

At the same time, the lattice model \eqref{eq:iS-Villain-lattice} can be viewed as a regularization of the continuum theory \eqref{eq:i action_continuum}, which is valid for all values of $\mu_0$ and $\mu$ (and correspondingly, all values of $\beta_0$ and $\beta$ as in \eqref{UKmumuzbeta}).  Therefore, the analysis of the lattice model \eqref{eq:iS-Villain-lattice} provides a rigorous justification to the continuum analysis of \eqref{eq:i action_continuum} in \cite{paper1}.

In Section \ref{sec:spectrum}, we will study the spectrum of the lattice model \eqref{eq:iS-Villain-lattice}.  We will formulate the lattice problem on a finite lattice with $L_x$ and $L_y$ sites in the $x$ and $y$ directions, respectively. For simplicity, we can take $L_x=L_y=L$.  Then, the spectrum includes three kinds of states with energies
\ie\label{latticespectrum}
&{\rm plane\ waves}\qquad \qquad &&{\sqrt{UK}\over L^2} ={1\over \sqrt{\mu_0 \mu} a^2 L^2} ~,\\
&{\rm momentum\ states}\qquad &&{U\over L}={1\over \mu_0 a^2L}~,\\
&{\rm winding\ states} \qquad \quad &&{K\over L}={1\over \mu a^2L}~.
\fe
Here, the momentum states and the winding states carry nonzero momentum and winding subsystem charges respectively.

For large $L$ with fixed $U$ and $K$, all these states go to zero energy, but as different powers of $L$.  The plane waves have parametrically lower energy.  Therefore, we can try to zoom on them and ignore the heavier momentum and winding states.

Section \ref{sec:cont} discusses the continuum field theory for the light plane waves.
In the continuum limit, we take $L_i\to \infty$ holding the physical lengths
\ie
\ell_i = {L_i\over a}
\fe
fixed.  In that limit, \eqref{latticespectrum} becomes
\ie\label{latticespectrumc}
&{\rm plane\ waves}\qquad \qquad &&{\sqrt{UK}\over L^2} ={1\over \sqrt{\mu_0 \mu} \ell^2} ~,\\
&{\rm momentum\ states}\qquad &&{U\over L}={1\over \mu_0 \ell a}~,\\
&{\rm winding\ states} \qquad \quad &&{K\over L}={1\over \mu  \ell a}~.
\fe
The plane waves have finite energy, while the energies of the momentum and winding states diverge as $1\over a$.  This is consistent with the findings in \cite{paper1}.  However, if we take $L\to\infty $ before taking the continuum limit, then all these states have zero energy.  This is a manifestation of the UV/IR mixing of our system  \cite{paper1}.

In Section \ref{sec:correlation}, we will study correlation functions of the lattice model \eqref{eq:iS-Villain-lattice}.    We will reproduce the continuum answers in \cite{paper1}, and will study more correlation functions and compare them with \cite{PhysRevB.66.054526}. Among other things, the analysis of the correlation functions will expose interesting consequences of the intricate UV/IR mixing in this theory.

One of the main points of our paper is the fact that the continuum Lagrangian \eqref{eq:i action_continuum} is an almost standard continuum field theory.  As a standard low-energy effective field theory, it is characterized by a scaling symmetry, \eqref{scalingxy}.\footnote{Usually, the phrases ``low-energy effective theory'' and ``long-distance effective theory'' are used interchangeably.  As we will discuss, because of the UV/IR mixing, this is not the case here.}  Also, it describes correctly the spectrum of plane waves and can be used to compute correlation functions of operators creating them.  There are however a number facts that make this theory ``nonstandard.''  First, as we mentioned above, these correlation functions have surprising UV singularities, which are regulated by an IR cutoff.  Second, when the theory is formulated on a slanted torus, it has a peculiar global symmetry and a strange ground state degeneracy \cite{Rudelius:2020kta}.  These two properties signal the UV/IR mixing of the theory.

Since the momentum and winding states are parametrically more energetic than the plane waves, the continuum field theory based on \eqref{eq:i action_continuum} does not include them as states.
We will also see that the lattice correlation functions of operators creating these states violate the scaling symmetry \eqref{scalingxy} of the continuum theory.  Indeed, these correlation functions vanish in the continuum limit. In the language of the renormalization group, we can interpret these operators as redundant operators.\footnote{Redundant operators are operators whose correlation functions decay exponentially on the scale of the UV cutoff and hence they vanish in the low energy theory.  Typical examples are operators that vanish because of the equations of motion and their correlation functions have only contact terms.  In our case, they are redundant for other reasons.}
This is consistent with the conclusions of \cite{paper1}, where we referred to this interpretation of the continuum theory as the ``conservative approach.''

Even though the continuum field theory \eqref{eq:i action_continuum} does not include the momentum and winding states, these states are still interesting because they exist in the regularized theory \eqref{eq:iS-Villain-lattice}.  Furthermore, they are the lightest states of the lattice system with nontrivial charges.  We can interpret the sectors of the theory built on top of these states as superselection sectors or as defects in the low-energy theory.\footnote{It is standard to have a field theory with a single Hilbert space that splits in the large volume limit to several distinct superselection sectors.  Here, the finite volume lattice system has a single Hilbert space and it splits in the continuum limit (in finite volume) to several superselection sectors labeled by the charges.  This situation is reminiscent of the case of the toric code \cite{Kitaev:1997wr}, where the Hilbert space of the finite volume lattice model has finite energy anyon excitations.   The low-energy theory is a continuum topological field theory.  The anyons are described as defects in that theory and the Hilbert space splits into different sectors.} The description of all these sectors was referred to in \cite{paper1} as the ``ambitious approach.''  Below we will discuss it in more detail.

In Section \ref{sec:dipole-deform}, we deform the lattice model \eqref{eq:iS-Villain-lattice} by the momentum dipole operators that break the $U(1)$ momentum subsystem symmetry to the ordinary (zero-form) $U(1)$ momentum symmetry. We show that there is a residual mixed 't Hooft anomaly between the ordinary momentum symmetry and the winding subsystem symmetry. It implies that the system remains gapless after the deformation. For small deformations, we expect the model to flow at low energies to the 2+1d $\phi$-theory \eqref{eq:i action_continuum}, whereas for large deformations, we expect it to flow to the ordinary 2+1d compact boson theory.

In Section \ref{sec:3+1d}, we will study the exotic 3+1d models of \cite{paper2}, again using their modified Villain version of \cite{Gorantla:2021svj}.  Unlike the 2+1d case, here we will encounter new subtleties in the way we take the low-energy limit and the continuum limit.
Specifically, the continuum limit is obtained not with fixed lattice couplings, but instead, we have to take them to zero or infinity appropriately.

Finally, in Section \ref{conclusions}, we summarize our main results.

In Appendix \ref{onepone}, we will review the continuum limit of the famous 1+1d XY model using the language of this paper.  This discussion is not new and it is presented here only for comparison with our analysis of the other models.

In Appendix \ref{sec:app_phidot2pt}, we study in great detail the two-point function of $\partial_\tau \phi$ in the lattice model \eqref{eq:iS-Villain-lattice}.   We present some exact expressions and consider interesting limits.

In Appendix \ref{sec:anomaly_lattice}, we provide more details on the analysis of the anomaly in the modified Villain model after turning on the momentum dipole deformation. We compute the anomaly explicitly by coupling the lattice model to lattice background gauge fields.

\section{2+1d XY-plaquette model and its modified Villain formulation}\label{sec:2+1d-XYplaq}

\subsection{2+1d XY-plaquette model}
In the Lagrangian formalism, the 2+1d XY-plaquette model \cite{PhysRevB.66.054526} is defined on a 2+1d Euclidean lattice with a phase $e^{i\phi}$ on each site.
 The   lattice coordinates are $\hat \tau =1,\cdots, L_\tau$, $\hat x=1,\cdots, L_x$, and $\hat y=1,\cdots, L_y$ with periodic boundary conditions.
 The Euclidean action is
\ie\label{eq:S_XY_plaq}
S=-b_0\sum_{\tau-\text{link}}\cos(\Delta_\tau \phi)-b\sum_{xy-\text{plaq}}\cos(\Delta_x\Delta_y\phi)~.
\fe
The phase diagram of this lattice model is controlled by the dimensionless coupling constants $b_0$ and $b$. The model has a $U(1)$ momentum subsystem symmetry that shifts $\phi\rightarrow\phi+f_x(\hat x)+f_y(\hat y)$.

The 2+1d XY-plaquette model in the Hamiltonian formalism is defined on a 2d spatial lattice. On each site $s$, there is a $U(1)$ variable $\phi_s$ and its conjugate momentum $\pi_s$ that obey $[\phi_s,\pi_{s'}]=i\delta_{s,s'}$. The Hamiltonian is
\ie\label{eq:H_XY_plaq}
H  = { \mathcal U \over2} \sum_{\text{site}} \pi^2 - {\mathcal K }\sum_{xy-\text{plaq}}\cos(\Delta_x\Delta_y\phi)~.
\fe
The coupling constants $\mathcal U$ and $\mathcal K$ both have mass dimension +1.
The charges of the momentum subsystem symmetry  are
\ie\label{eq:momentum_charge}
N_x(\hat x)=\sum_{\hat y}\pi_{\hat x,\hat y}~,\quad  N_y(\hat y)=\sum_{\hat x}\pi_{\hat x,\hat y}~.
\fe

The phase diagram of this model is controlled by the dimensionless parameter $\mathcal{U}/\mathcal{K}$.
When $\mathcal{U}/\mathcal{K}$ is large, the low energy phase is gapped and when $\mathcal{U}/\mathcal{K}$ is small it is gapless \cite{PhysRevB.66.054526}.

\subsection{Modified Villain formulation}\label{sec:mV}

In \cite{Gorantla:2021svj}, a modified Villain version of the XY-plaquette model was introduced:
\ie\label{eq:S-Villain-lattice}
S=\frac{\beta_0}{2}\sum_{\tau-\text{link}}(\Delta_\tau\phi-2\pi n_\tau)^2+\frac{\beta}{2}\sum_{xy-\text{plaq}}(\Delta_x\Delta_y\phi-2\pi n_{xy})^2+i\sum_{\text{cube}}\phi^{xy}(\Delta_\tau n_{xy}-\Delta_x\Delta_y n_\tau)~,
\fe
where $\phi^{xy}$ is a $\mathbb{R}$-valued field on the cubes and $(n_\tau,n_{xy})$ are $\mathbb{Z}$-valued tensor gauge fields on $\tau$-links and $xy$-plaquettes. The theory has $\mathbb{Z}$ gauge symmetries
\ie
&\phi\sim\phi + 2\pi k~,
\\
&\phi^{xy}\sim \phi^{xy}+2\pi k^{xy}~.
\\
&(n_\tau,n_{xy})\sim (n_\tau+\Delta_\tau k,n_{xy}+\Delta_x\Delta_y k)~,
\fe
which make the fields $\phi$ and $\phi^{xy}$ compact.  For $\beta_0,\beta \gg 1$, this model is the same as the model based on \eqref{eq:S_XY_plaq} with $\beta\approx b$ and $\beta_0 \approx b_0$.  In the rest of this paper, we will be interested in the model \eqref{eq:S-Villain-lattice} with generic $\beta$ and $\beta_0$, which is different than the model based on \eqref{eq:S_XY_plaq}.

\bigskip\centerline{\it Global symmetries, anomalies, and dualities}\bigskip

The modified Villain model is very close to the continuum field theory (which we will be discussed in Section \ref{sec:cont}).
It realizes the exact $U(1)$ momentum and $U(1)$ winding subsystem symmetries that shift $\phi\rightarrow\phi+f_x(\hat x)+f_y(\hat y)$ and $\phi^{xy}\rightarrow\phi^{xy}+f^{xy}_x(\hat x)+f^{xy}_y(\hat y)$ respectively.

Furthermore, there is a mixed 't Hooft anomaly between these two $U(1)$ symmetries \cite{Gorantla:2021svj}.  To see that, we couple these symmetries to classical background fields $(A_\tau,A_{xy})$ and $(\tilde A^{xy}_\tau,\tilde A)$ with gauge transformation parameters $\alpha$ and $\tilde \alpha^{xy}$ respectively.  In Appendix \ref{sec:anomaly_lattice}, we discuss the anomaly on the lattice.  In the continuum, the anomalous gauge variation is given by
\ie\label{XYplaq-anomaly-lattice-continuum}
\frac{i}{2\pi}\int d\tau dx dy\,\alpha (\d_\tau \tilde A - \d_x \d_y \tilde A_\tau^{xy} ) ~.
\fe
It signals an 't Hooft anomaly because it cannot be canceled by any 2+1d local counterterms.

The existence of this anomaly implies that the modified Villain model has to be gapless for all values of its coupling constants. As discussed in Section \ref{sec:cont}, the low-energy theory of this lattice model is described by the gapless continuum $\phi$-theory \eqref{eq:action_continuum}.

This points to an important difference between the XY-plaquette model and this modified Villain model.  For large $\mathcal K\gg \mathcal U$ they are the same.  Both of them are gapless.  However, for $\mathcal K\ll \mathcal U$ the XY-plaquette model is gapped \cite{PhysRevB.66.054526}, while the modified Villain theory is still gapless.

Applying the Poisson resummation on the $\mathbb{Z}$-valued tensor gauge fields, the lattice model has an exact self-duality that maps the couplings $(\beta_0,\beta)$ to $({1}/{4\pi^2\beta},{1}/{4\pi^2\beta_0})$. The self-duality swaps the momentum and winding subsystem symmetries \cite{Gorantla:2021svj}.

\bigskip\centerline{\it Hamiltonian formalism}\bigskip

The modified Villain model can also be formulated in the Hamiltonian formalism. We will start with a Eucldiean lattice of size $L_x$, $L_y$, $L_\tau$.

First, we integrate out $\phi^{xy}$, which enforces the $\mathbb{Z}$-valued gauge fields to be flat.
We  then  gauge fix most of the $n_\tau(\hat \tau,\hat x,\hat y)$, $n_{xy}(\hat \tau,\hat x,\hat y)$ to zero, except for
\ie
&n_{\tau}(0,\hat x,\hat y)\equiv \bar n^x(\hat x)+\bar n^y(\hat y)~,
\\
&n_{xy}(\hat\tau,\hat x,0)\equiv\bar n^{xy}_y(\hat x)~,
\\
&n_{xy}(\hat\tau,0,\hat y)\equiv \bar n^{xy}_y(\hat y)~.
\fe
The latter  are the gauge-invariant holonomies. At $\hat x =\hat y=0$, we have $\bar n^{xy}_x(0)=\bar n^{xy}_y(0)$. This gauge choice leaves a residual time-independent gauge symmetry
\ie\label{eq:residue_gauge}
\phi(\hat\tau,\hat x,\hat y)\sim \phi(\hat \tau,\hat x,\hat y)+2\pi k_x(\hat x)+2\pi k_y(\hat y)~,\qquad k_x(\hat x), k_y(\hat y) \in \mathbb{Z}~.
\fe
Imposing invariance under this gauge symmetry can be thought of as imposing Gauss law of the tensor $\mathbb{Z}$ gauge symmetry.
 Equivalently, the sum over $\bar n^x (\hat x), \bar n^y (\hat y)$
makes the momentum modes of $\phi$ compact and thus quantizes their conjugate momenta.

Following \cite{Gorantla:2021svj}, we can define a new field $\bar\phi$ on the sites such that in the fundamental domain $\bar \phi(\hat \tau,\hat x,\hat y) = \phi(\hat \tau,\hat x,\hat y)$ for $1\leq \hat x^\mu \leq L_\mu$, and beyond the fundamental domain, it is extended via
\ie\label{XYplaq-phibar}
&\bar \phi(\hat \tau+L_\tau,\hat x,\hat y) = \bar \phi(\hat \tau,\hat x,\hat y) - 2\pi \bar n^x(\hat x) - 2\pi \bar n^y(\hat y)~,
\\
&\bar \phi(\hat \tau,\hat x+L_x,\hat y) = \bar \phi(\hat \tau,\hat x,\hat y) - 2\pi \sum_{\hat y'=0}^{\hat y-1}\bar n^{xy}_y(\hat y')~,
\\
&\bar \phi(\hat \tau,\hat x,\hat y+L_y) = \bar \phi(\hat \tau,\hat x,\hat y) - 2\pi \sum_{\hat x'=0}^{\hat x-1}\bar n^{xy}_x(\hat x')~.
\fe
The new field $\bar{\phi}$ satisfies $\Delta_\tau \bar \phi = \Delta_\tau \phi - 2\pi n_\tau$, and  $\Delta_x \Delta_y \bar \phi = \Delta_x \Delta_y \phi - 2\pi n_{xy}$ in the above gauge choice. Although $\phi$ is single-valued, $\bar \phi$ can wind around the nontrivial cycles of spacetime. In the path integral, we should sum over nontrivial winding sectors of $\bar\phi$. The action in terms of $\bar \phi$ is
\ie
&\frac{\beta_0}{2} \sum_{\tau\text{-link}} (\Delta_\tau \bar\phi )^2 + \frac{\beta}{2} \sum_{xy\text{-plaq}} (\Delta_x \Delta_y \bar\phi )^2  ~,
\fe

Next, we make the time non-compact, introduce a continuous time $\tau=a_\tau \hat \tau$ and take the following limit
\ie\label{relatepara}
a_\tau\rightarrow 0~, \quad U\equiv \frac{1}{\beta_0 a_\tau}=\text{fixed}~,\quad K\equiv\frac{\beta}{a_\tau}=\text{fixed}~.
\fe
The Euclidean action becomes
\ie\label{eq:S_lattice_villain}
S&=\int d\tau\left[ \frac{1}{2U}\sum_{\text{site}}(\partial_\tau\bar\phi)^2+\frac{K}{2}\sum_{xy-\text{plaq}}(\Delta_x\Delta_y\bar\phi)^2\right]
\\
&=\int d\tau\left\{ \frac{1}{2U}\sum_{\text{site}}(\partial_\tau\phi)^2+\frac{K}{2}\sum_{xy-\text{plaq}}\Big[\Delta_x\Delta_y\phi-2\pi \big(\bar n^{xy}_x(\hat x)\delta_{\hat y,0}+\bar n^{xy}_y(\hat y)\delta_{\hat x,0}(1-\delta_{\hat y,0})\big)\Big]^2\right\}
~.
\fe

The Hamiltonian is
\ie\label{eq:H_lattice_villan}
H  = { U \over2} \sum_{\text{site}} \pi^2 + \frac{K}{2}\sum_{xy-\text{plaq}}\Big[\Delta_x\Delta_y\phi-2\pi \big(W_x(\hat x)\delta_{\hat y,0}+W_y(\hat y)\delta_{\hat x,0}-W\delta_{\hat x,0}\delta_{\hat y,0}\big)\Big]^2\,.
\fe
where the $W$'s are the winding charges
\ie
&W_x(\hat x)=\sum_{\hat y} \Delta_x\Delta_y\bar\phi(\hat \tau,\hat x,\hat y)=(1-\delta_{\hat x,0})\bar n^{xy}_x(\hat x)+\delta_{\hat x,0}\sum_{\hat y}\bar n^{xy}_y(\hat y)~,
\\
&W_y(\hat y)=\sum_{\hat x} \Delta_x\Delta_y\bar\phi(\hat \tau,\hat x,\hat y)=(1-\delta_{\hat y,0})\bar n^{xy}_y(\hat y)+\delta_{\hat y,0}\sum_{\hat x}\bar n^{xy}_y(\hat x)~,
\\
&W = \sum_{\hat x} W_x(\hat x)=\sum_{\hat y} W_y(\hat y)~.
\fe
The momentum charges are the same as \eqref{eq:momentum_charge}.

Finally, in the Hamiltonian formalism, the wave functions depend on the values of the real field $\phi$ and the integer fields $\bar n_{x}^{xy}(\hat x)$, $\bar n_{y}^{xy}(\hat y)$.
Because of the gauge symmetry \eqref{eq:residue_gauge}, the values of the field $\phi$ at different spatial points $\hat x,\hat y$ are not independent.
Therefore, the Hilbert space is not quite a tensor product of local Hilbert spaces at every site.  This fact makes it possible to have an 't Hooft anomaly in the system.

\section{The spectrum}\label{sec:spectrum}

In this section, we examine the spectrum of the lattice model \eqref{eq:H_lattice_villan}.
We will then discuss the robustness of these field theories under deformations that break the $U(1)$  subsystem symmetries.
We place the lattice model on a 2d spatial lattice of size $L_x$, $L_y$ with periodic boundary condition.

The discussion of the spectrum of the modified Villain lattice model proceeds similarly as that for the continuum field theory \cite{paper1}.
The spectrum can be organized according to the global momentum and winding subsystem symmetries:
\begin{itemize}
	\item
	States that are not charged under any subsystem symmetries. These states are plane waves with quantized (spatial) momenta
	\ie
	\phi(t,\hat x,\hat y)=\sum_{n_x=1}^{L_x-1}\sum_{n_y=1}^{L_y-1}C_{n_x,n_y} \exp\left(i\omega_{n_x,n_y} t+i\frac{2\pi  n_x}{L_x} \hat x +i\frac{2\pi  n_y}{L_y} \hat y\right)~.
	\fe
	The dispersion relation is
	\ie
	\omega_{n_x,n_y} = 4\sqrt{UK} \left|\sin {n_x \pi\over L_x} \sin {n_y \pi\over L_y} \right|~.
	\fe
	Here we exclude the plane waves with either $n_x=0$ or $n_y=0$. We will discuss them soon.
	
	For large $L_x\sim L_y\sim L$, the plane waves can be divided into three classes: generic plane waves with $n_i$ of order $L$ have energies of order $\sqrt{UK}$; the plane waves with  $n_x$ of order one but  $n_y$ of order $L$ have energies of order $\sqrt{UK}/L$ (and similarly with $x,y$ exchanged); the plane waves with  $n_x,n_y$ both of order one have energies of order ${\sqrt{UK}}/{L^2}$. The highest energy these plane waves can have is $4\sqrt{UK}$.
	\item
	States that are charged under   the momentum subsystem symmetry. We will refer to these states as momentum states. Classically, the momentum subsystem symmetry maps plane waves with zero-energy to each other so the momentum states have zero classical energy. However, because of the identification \eqref{eq:residue_gauge}, these momentum states are lifted upon quantization and their energies are determined by their quantized momentum charges. A momentum state with momentum charges $N_x(\hat x)$, $N_y(\hat y)$ has energy
	\ie\label{XYmomentum}
	E=\frac{U}{2 L_x L_y}\left[L_x\sum_{\hat x} N_x(\hat x)^2+L_x\sum_{\hat y} N_y(\hat y)^2-N^2\right]~,
	\fe
	where $N = \sum_{\hat x} N_x(\hat x)=\sum_{\hat y} N_y(\hat y)$.
	
	For large $L_x\sim L_y\sim L$, these momentum states have energies of order ${U}/{L}$.
	
	\item
	States that are charged under  the winding subsystem symmetry. We will refer to these states as winding states. The minimal energy winding configuration with winding charges $W_x(\hat x)$, $W_y(\hat y)$ takes the form\footnote{The naive winding configurations that are simply linear in $\hat x$ or $\hat y$ are part of the momentum states \eqref{XYmomentum} and as such they are lifted.  Equivalently, these states have vanishing $U(1)$ winding subsystem symmetry charge \cite{paper1,Rudelius:2020kta}. }
	\ie
	\phi=2\pi\left[\frac{\hat x}{L_x}\left(\sum_{\hat y_0}W_y(\hat y_0)\Theta(\hat y-\hat y_0)\right)+\frac{\hat y}{L_y}\left(\sum_{\hat x_0}W_x(\hat x_0)\Theta(\hat x-\hat x_0)\right)-W\frac{\hat x\hat y}{L_xL_y}\right]
	\fe
	for $1\leq\hat x\leq L_x$, $1\leq\hat y\leq L_y$ where $W = \sum_{\hat x} W_x(\hat x)=\sum_{\hat y} W_y(\hat y)$.
	Its energy is
	\ie
	E=\frac{2\pi^2 K}{L_xL_y}\left[L_x\sum_{\hat x_0}W^x(\hat x_0)^2+L_y\sum_{\hat y_0}W^y(\hat y_0)^2-W^2\right]~.
	\fe
	
	For large $L_x\sim L_y\sim L$, these winding states have energies of order ${K}/{L}$.
	\item
	States that are charged under both the momentum and winding subsystem symmetries.
	
	For large $L_x\sim L_y\sim L$, these states have energies of order $\sqrt{UK}/L$ up to a function of $U/K$.
\end{itemize}

\begin{table}
\begin{center}
\begin{tabular}{|c|c|c|c|}
\hline
Lattice model & $E_\text{wave}$ & $E_\text{mom/elec}$ & $E_\text{wind/mag}$
\tabularnewline
\hline
1+1d XY model & $\frac{1}{L}$ & $\frac{1}{L}$ & $\frac{1}{L}$
\tabularnewline
\hline
2+1d XY-plaquette model & $\frac{1}{L^2}$ & $\frac{1}{L}$ & $\frac{1}{L}$
\tabularnewline
\hline
\end{tabular}
\caption{The $L$ dependence of the energies of the three kinds of the states in (the modified Villain formulation of) the 2+1d XY-plaquette model \cite{Gorantla:2021svj}.
We compare these scalings with those in  (the modified Villain formulation of) the ordinary 1+1d  XY model,   which are reviewed in Appendix \ref{onepone}.
Since the gapless phase of the latter is described by a conformal field theory (CFT), all the low-lying states scale to zero as $1/L$.
}  \label{tbl:lattice-energythr}
\end{center}
\end{table}

We summarize that the low-lying plane waves (charge neutral states) have energies of order $1/L^2$, the charged states have minimal energies of order $1/L$ and generic states have energies of order $1$.  See  Table \ref{tbl:lattice-energythr} where, for comparison, we included also the analogous expressions for the states in the (modified Villain version of the) 1+1d XY model.
(See Appendix \ref{onepone} for a brief review of the ordinary 1+1d XY model.)
 We see that our system is special because, as $L\to \infty$, the plane waves are parametrically lighter than the charged states.  This fact has important consequences, which we will discuss below.

\section{Continuum $\phi$-theory}\label{sec:cont}

Our goal here is to find a low-energy continuum field theory for this lattice theory.  We start from the spectrum we worked out above and scale the parameters such that only the low-lying states survive.  Specifically, since the plane-waves have the lowest energy, we should zoom on them and construct the theory as follows.

We start with the Euclidean action \eqref{eq:S_lattice_villain} and introduce continuous coordinates $x=a\hat x$, $y=a\hat y$.  Then, we take the continuum limit
\ie\label{eq:continuum_limit}
a\rightarrow 0~, \quad \mu_0=\frac{1}{a^2U}=\text{fixed}~,\quad \mu=\frac{1}{a^2K}=\text{fixed}~,\quad \ell_x=aL_x=\text{fixed}~,\quad \ell_y=aL_y=\text{fixed}~.
\fe
Here $\ell_x$ and $\ell_y$ are the physical sizes of the system in the $x$ and $y$ directions.

This leads to the continuum field theory with the Euclidean action
\ie\label{eq:action_continuum}
S=\int d\tau dx dy\left[\frac{\mu_0}{2}(\partial_\tau\phi)^2+\frac{1}{2\mu}(\partial_x\partial_y\phi)^2\right]~,
\fe
where the continuum field $\phi$ comes from $\bar\phi$ on the lattice \eqref{XYplaq-phibar}. (For simplicity, we drop the bar over $\phi$.) The field $\phi$ is subject to an identification:
\ie\label{nxyid}
\phi(\tau,x,y)\sim\phi(\tau,x,y)+2\pi n_x(x)+2\pi n_y(y)~,\qquad n_x,n_y\in\mathbb{Z}.
\fe
This gauge symmetry is the continuum counterpart of the gauge symmetry of our lattice discussion, which leads to the winding sectors of $\phi$ and to the Gauss law constraint of \eqref{eq:residue_gauge}.
This is the Euclidean version of the 2+1d $\phi$-theory of \cite{paper1}, which had been first introduced in \cite{PhysRevB.66.054526}.
(See also \cite{Tay_2011,You:2019cvs,You:2019bvu,Karch:2020yuy,You:2021tmm}  for related discussions on this theory.)

Inherited from the modified Villain lattice model, the continuum $\phi$-theory has $U(1)$ momentum and winding subsystem symmetries with a mixed 't Hooft anomaly \eqref{XYplaq-anomaly-lattice-continuum} \cite{Gorantla:2021svj}, and an exact self-duality that maps the theory with coupling $(\mu_0,\mu)$ to the theory with coupling $({\mu}/{4\pi^2},4\pi^2\mu_0)$ \cite{paper1}.

The spectrum of this continuum theory can be derived by using the scaling \eqref{eq:continuum_limit} in the expressions for the spectrum of the lattice theory.  The energies of the low-lying plane waves (charge neutral states) are of order $1/ \ell^2$, the minimal energies of the charged states are of order $1/ \ell a$, and the energies of generic states are of order $1/ a^2$. This reproduces the hierarchy of the spectrum in \cite{paper1}.  We summarize it in Table \ref{tbl:cont2-energyth}, where we also include, for comparison, a non-exotic 1+1d model.

\begin{table}[t]
\begin{center}
\begin{tabular}{|c|c|c|c|c|c|}
\hline
Lattice model & Continuum Theory & $E_\text{wave}$ & $E_\text{mom/elec}$ & $E_\text{wind/mag}$ & $a_\tau$
\tabularnewline
\hline
1+1d XY model& 1+1d compact boson & $\frac{1}{\ell}$ & $\frac{1}{\ell}$ & $\frac{1}{\ell}$ & $a$
\tabularnewline
\hline
2+1d XY-plaquette model& 2+1d $\phi$-theory & $\frac{1}{\ell^2}$ & $\frac{1}{a \ell}$ & $\frac{1}{a \ell}$ & $a^2$
\tabularnewline
\hline
\end{tabular}
\caption{The $\ell $ dependence of the energies of the three kinds of states in the continuum limit $a \rightarrow 0$ of the models in Table \ref{tbl:lattice-energythr}.  Here we suppressed dimensionful coefficients, which are finite in the continuum limit. We also included the scaling of $a_\tau$ in terms of $a$, such that the energies of the plane waves are kept finite.  Note that in the ordinary 1+1d XY model, $a_\tau \sim a$, while in the 2+1d XY-plaquette model $a_\tau \sim a^2$.  This is related to zooming on the states with energy of order $1/L^2$ and pushing the states with energy of order $1/L$ to infinity.  Also, comparing with \eqref{UKmumuzbeta}, we see that lattice coupling constants $\beta_0$ and $\beta$ are not scaled in the continuum limit.  The same is true in the 1+1d XY model.}\label{tbl:cont2-energyth}
\end{center}
\end{table}

The charged states are infinitely heavier than the plane waves.  Therefore, it is natural to ignore them in the continuum limit and conclude that the spectrum of the continuum theory includes only the plane wave states.

This can also be phrased as follows.  In a Euclidean formulation, we limit ourselves to configurations with finite action.  This removes the winding states.  In order to remove the momentum states we replace the action \eqref{eq:action_continuum} by
\ie\label{Euclfor}
S=\int d\tau dx dy\left[\frac{\mu_0}{2}\left(\partial_\tau\phi(x,y,\tau)-A^{(x)}_\tau(x,\tau) -A^{(y)}_\tau(y,\tau)\right)^2+\frac{1}{2\mu}\left(\partial_x\partial_y\phi(x,y,\tau)\right)^2\right]~,
\fe
which has the gauge symmetry
\ie\label{gaugealphal}
&\phi(x,y,\tau)\to \phi(x,y,\tau)+\alpha^{(x)}(x,\tau)+\alpha^{(y)}(y,\tau),\\
&A^{(x)}_\tau(x,\tau) \to A^{(x)}_\tau(x,\tau) +\partial_\tau \alpha^{(x)}(x,\tau)+\lambda_\tau(\tau),\\
&A^{(y)}_\tau(y,\tau) \to A^{(y)}_\tau(y,\tau)+\partial_\tau \alpha^{(y)}(y,\tau)-\lambda_\tau(\tau)~.
\fe
In the temporal gauge $A^{(x)}_\tau(x,\tau) =A^{(y)}_\tau(y,\tau)=0$, we recover the original action \eqref{eq:action_continuum}, but Gauss law states that the momentum charges vanish.

Equivalently, in a Hamiltonian formulation our states are associated with wave functional of the configuration space.  Since our space is a two-torus, the coordinates are the set ${\cal C}=\{\phi(x, y)\}$ of maps from the two-torus parameterized by $x$ and $y$ to a circle parameterized by $\phi \sim \phi+2\pi$. We impose that the map has finite $\partial_x\partial_y\phi$.  This eliminates the winding states.  The momentum states are removed by imposing an identification on $\cal C$.  Let $f_x(x)\in {\cal C}$ and $f_y(y)\in {\cal C}$ be maps from the torus to the circle that depend only on $x$ or only on $y$. Then, our configuration space is the quotient of $\cal C$ by the identification
\ie\label{calCid}
\phi(x,y)\sim \phi(x,y)+f_x(x)+f_y(y)~.
\fe
This identification is the Gauss law constraint of the Euclidean formulation in \eqref{Euclfor} and it sets the momentum charges to zero.

Actually, our low-energy theory is richer than we have just stated.  Above, we considered space to be a two-torus with purely imaginary modular parameter $\tau =\tau_1+i\tau_2$ in our $(x,y)$ coordinate system.  However, we can study our lattice theory or the continuum theory on a slanted torus with nonzero $\tau_1$ in that coordinate system. For rational $\tau_1 ={k\over m}$, the system has a subtle  $\mathbb{Z}_m\times \mathbb{Z}_m$ momentum and winding symmetry, which is realized projectively on the Hilbert space.  This leads to an $m$-fold degeneracy in the spectrum \cite{Rudelius:2020kta}, and demonstrates that the low-energy theory does depend on the underlying subsystem symmetry.\footnote{For infinite $L$, $\tau_1$ can be irrational and then the ground state is infinitely degenerate.}

Our low-energy theory provides a universal  description for both the modified Villain model \eqref{eq:S_lattice_villain}, \eqref{eq:H_lattice_villan}, as well as the original models \eqref{eq:S_XY_plaq}, \eqref{eq:H_XY_plaq} in the gapless phase, at  energies comparable to $1/L^2$.  It is not affected by deforming the theory \eqref{eq:H_lattice_villan} by terms  preserving the momentum and winding subsystem symmetries. This reflects the universality of this low-energy theory \cite{paper1}.

Following \cite{paper1}, we can also study the heavier momentum and winding states using this continuum field theory.  These states are infinitely heavier than the plane waves and therefore they are not states in the Hilbert space of the continuum theory.  However, since they carry conserved charges and since there are no lighter states with the same charges, they can be thought of as superselection sectors or defects in our continuum theory. The winding states are treated as discontinuous fields with infinite energy \cite{paper1}.  The momentum states are  defects of the form $e^{i\int d\tau \left(A^{(x)}_\tau(x,\tau) + A^{(y)}_\tau(y,\tau)\right)}$, which are invariant under the gauge symmetry \eqref{gaugealphal}.  This modifies the Gauss law constraints such that the states are not invariant under the operator implementing the identification \eqref{calCid} -- they are charged under it \cite{paper1}.

\section{Correlation functions}\label{sec:correlation}

In this section, we compute several correlation functions in the modified Villain version of the 2+1d XY-plaquette model with finite number of sites $L_x$ and $L_y$. We want to study the behavior of these correlation functions in various limits, and under various scalings.
Different limits of the correlation functions will probe different sets of states in Section \ref{sec:spectrum}.

Ignoring the integer gauge fields, the Hamiltonian is given by \eqref{eq:H_lattice_villan}
\ie\label{eq:H_lattice_villan2}
H  = {U \over2} \sum_{\hat x=1}^{L_x} \sum_{\hat y=1}^{L_y} \pi_{\hat x,\hat y}^2 + {K \over 2}\sum_{\hat x=1}^{L_x}\sum_{\hat y=1}^{L_y} (\Delta_x \Delta_y\phi_{\hat x,\hat y})^2~,
\fe
where $\pi_{\hat x, \hat y}$ and $\phi_{\hat x,\hat y}$ obey $[\phi_{\hat x,\hat y} , \pi_{\hat x',\hat y'} ] = i \delta_{\hat x, \hat x'}\delta_{\hat y,\hat y'}$.

Alternatively, we can view this Hamiltonian as the one derived from the original 2+1d XY-plaquette model \eqref{eq:H_XY_plaq} in the ${\cal K}\gg {\cal U}$ limit.
In this limit, the model is gapless and the Hamiltonian can be approximated by \eqref{eq:H_lattice_villan2} with $K\approx {\cal K}$ and $U\approx {\cal U}$.

We will compute correlation functions of operators built out of $\phi$.
Using the exact duality of the modified Villain model (see Section \ref{sec:mV}), the correlation functions of operators built out of $\phi^{xy}$ can be obtained from those of $\phi$ by $(U,K)\rightarrow (4\pi^2K, U/4\pi^2)$.

Let $F(\hat x , \hat y, \tau)$ be minus the lattice propagator of $\phi$:\footnote{For simplicity, we assume both $L_x$ and $L_y$ are even so that $L_i/2$ in the range of the sum is an integer. We will also assume $\hat x,\hat y\ge 0$.}
\ie\label{latpro}
&F(\hat x , \hat y,\tau)= - \langle \phi_{\hat x,\hat y}(\tau)\phi_{0,0}(0)\rangle= F^{(0)}(\hat x , \hat y,\tau)+F^{(1)}(\hat x , \hat y,\tau)~,\\
&F^{(0)}(\hat x , \hat y,\tau)= { U \over2 L_xL_y} \left(L_x \delta_{\hat x , 0}+L_y \delta_{\hat y,0}-1\right)|\tau|~,\\
&F^{(1)}(\hat x , \hat y,\tau)\\
&={U\over 2L_xL_y }  \sum_{-{L_x\over 2} <n_x\le {L_x\over2} \atop n_x\neq 0}
\sum_{-{L_y\over2}<n_y\le {L_y\over2} \atop n_y\neq0}  {\cos({2\pi n_x \hat x\over L_x})+\cos({2\pi n_y \hat y\over L_y})-1- e^{-\omega_{n_x,n_y}| \tau|}\cos(\frac{2\pi  n_x\hat x}{L_x})\cos(\frac{2\pi  n_y\hat y}{L_y})\over \omega_{n_x,n_y}}~.
\fe
Here, $F^{(0)}(\hat x,\hat y,\tau)$ arises from terms with $n_x=0$ or $n_y=0$.
Because of the subsystem symmetry, shifting the propagator by a function of $\hat x$ plus a function of $\hat y$ does not affect any physical correlation function. We used this freedom to choose the above form of $F^{(1)}(\hat x , \hat y,\tau)$ so that when we replace the sums over $n_i$ by integrals, there will not be any IR divergences.

In addition to $U/K$ and $L_i$, we have three dimensionless parameters: $\hat x$, $\hat y$, and $T\equiv 4\sqrt{UK}|\tau|$. We will study the behaviour of several correlation functions in different limits of these parameters, especially $L_i$ and $T$.

More specifically, starting from the lattice model on a finite rectangular lattice, we will consider the following limits to reproduce the results in \cite{PhysRevB.66.054526,paper1}:
\begin{itemize}
\item Continuum limit:
\ie\label{contlimit}
&1 \ll T \sim L_xL_y\,,~~~~\hat x^i \sim L_i\,,~~~~ U/K=\text{fixed}\,.
\fe
 To understand this limit better, we write $U,K,L_i, T,\hat x^i$ in terms of the continuum variables $\mu_0,\mu,\ell_i,\tau, x^i$:
\ie\label{contvariable}
U = {1\over \mu_0 a^2}\,,~~~K = {1\over \mu a^2} \,,~~~
L_i  ={ \ell_i\over a}\,,~~~~T=  {4\over \sqrt{\mu\mu_0}} {|\tau|\over a^2}\,,~~~~\hat x^i = {x^i\over a}\,.
\fe
This   limit is then equivalent to taking $a\to0$, while holding the other continuum variables (including the size $\ell_i$ of the system) fixed.
In this limit, the correlation functions receive contributions only from  the states with energy of order $1/L^2$,  discussed in Section \ref{sec:spectrum}.
This is the continuum limit in \cite{paper1}, and we will reproduce the correlation functions computed there.

\item Thermodynamic limit:
\ie\label{thermolimit}
&1 \sim T ,\hat x^i \ll L_x,L_y\,.
\fe
In terms of the continuum variables, this limit corresponds to taking $\ell_i\to\infty$ first, while holding the other variables including the lattice spacing $a$ fixed.
In this limit, the correlation functions receive contributions from both the states with energy of order $1/L^2$ and  $1/L$, discussed in Section \ref{sec:spectrum}.
For some of the correlation functions, we will subsequently take the $T\gg 1$ limit and recover the correlation functions in \cite{PhysRevB.66.054526}.
\end{itemize}

\subsection{$\langle \partial_\tau \phi \partial_\tau \phi \rangle$}\label{sec:phidot}

In this subsection we will always assume $|\tau|\neq0$.
The two-point function of $\partial_\tau \phi$ is:\footnote{ Using the exact duality in section \ref{sec:mV}, we can obtain the two-point function of $\Delta_x\Delta_y \phi$ from the two point function of $\partial_\tau \phi$ by $(U,K)\rightarrow(4\pi^2K,U/4\pi^2)$ with an additional overall factor $1/4\pi^2K$. }
\ie\label{phidot-2pt}
&\langle \partial_\tau \phi_{\hat x,\hat y}(\tau) \partial_\tau \phi_{0,0}(0)\rangle = \partial_\tau^2 F^{(1)}(\hat x, \hat y, \tau)
\\
& = -{2UK\over L_xL_y }\sqrt{U\over K}  \\
&\times\sum_{-{L_x\over 2} <n_x\le {L_x\over2} \atop n_x\neq 0} \sum_{-{L_y\over 2} <n_y\le {L_y\over2} \atop n_y\neq 0} \left| \sin\left({\pi n_x\over L_x}\right) \sin\left({\pi n_y\over L_y}\right)\right|\,
e^{-|\sin({\pi n_x\over L_x})\sin({\pi n_y\over L_y})| T}\cos\left(\frac{2\pi  n_x\hat x}{L_x}\right)\cos\left(\frac{2\pi  n_y\hat y}{L_y}\right)~.
\fe

\subsubsection{Continuum limit }

In the continuum limit \eqref{contlimit}, the two-point function becomes\footnote{Here, in the limit $L_i\rightarrow \infty$, we write $\sum_{-{L_i\over 2} <n_i\le {L_i\over2} \atop n_i\neq 0}\rightarrow 2 \sum_{n_i = 1}^\infty$.}
\ie\label{phidot-2pt-A.5}
\langle \partial_\tau\phi_{\hat x,\hat y}(\tau) \partial_\tau \phi_{0,0}(0)\rangle
\to -{8\pi^2 \over \mu^{1\over2}\mu_0^{3\over2}\ell_x^2 \ell_y^2 }  \sum_{n_x=1}^{\infty}\sum_{n_y=1}^{\infty} n_x n_y e^{-{4\pi^2\over \sqrt{\mu\mu_0}} {|\tau|\over \ell_x\ell_y} n_xn_y  }\cos\left(2\pi  n_x {x\over \ell_x}\right)\cos\left(2\pi  n_y{ y\over \ell_y}\right)~.
\fe
The sum is evaluated in various limits in Appendix \ref{sec:app_continuum2pt}. We summarize them below:
\ie\label{dotphi2pt}
&\langle \partial_\tau \phi_{\hat x,\hat y}(\tau) \partial_\tau\phi_{0,0}(0)\rangle \to- {1\over  2\pi^2 } {\sqrt{\mu\over \mu_0} } {1\over \tau^2}G\left({xy\sqrt{\mu\mu_0}\over |\tau|},{x\over \ell_x}, {y\over \ell_y} \right)
\\
& G\left({xy\sqrt{\mu\mu_0}\over |\tau|},{x\over \ell_x}, {y\over \ell_y} \right)=
\begin{cases}
 \dfrac{\tau^2 }{ (xy)^2 \mu\mu_0} & {|\tau |\over\sqrt{\mu\mu_0} }\ll xy\ ,\ x\ll \ell_x\ , \ y\ll \ell_y
 \\
 \log \left(\dfrac{|\tau|}{ xy\sqrt{\mu\mu_0}}\right) &  xy\ll {|\tau|\over \sqrt{\mu\mu_0}}\ , \ {|\tau|\over \ell_y \sqrt{\mu\mu_0}} \ll x\ , \ {|\tau|\over \ell_x \sqrt{\mu\mu_0}} \ll y
 \\
 \log\left(\dfrac{\ell_y}{ 2\pi y}\right) & x\ll {|\tau|\over \ell_y \sqrt{\mu\mu_0}} \ ,\  {|\tau|\over \ell_x \sqrt{\mu\mu_0}} \ll y\ll \ell_y
 \\
 \log\left(\dfrac{\ell_x}{ 2\pi x}\right) & y\ll {|\tau|\over \ell_x \sqrt{\mu\mu_0}} \ ,\  {|\tau|\over \ell_y \sqrt{\mu\mu_0}} \ll x\ll \ell_x
 \\
 \log\left( \dfrac{\ell_x\ell_y\sqrt{\mu\mu_0} }{ 4\pi^2 |\tau|}\right)
  &x\ll {|\tau|\over \ell_y \sqrt{\mu\mu_0}} \ ,\   y\ll {|\tau|\over \ell_x \sqrt{\mu\mu_0}}\ , \ \ell_x\ell_y\gg{ |\tau|\over \sqrt{\mu\mu_0}}
\end{cases}
\fe

As is clear from the first line in \eqref{dotphi2pt}, this correlation function has two scale symmetries:
\ie
\tau\rightarrow\lambda_x\lambda_y\tau,\quad x\rightarrow \lambda_x x,\quad y\rightarrow \lambda_y y,\quad \ell_x\rightarrow \lambda_x\ell_x~,\quad \ell_y\rightarrow \lambda_y\ell_y~.
\fe
Under both scale transformations, $\partial_\tau \phi$ has scaling dimension $\Delta=1$.  This is consistent with the scaling symmetry \eqref{scalingxy} of the continuum action.

We see that this correlation function has a surprising singularity at nonzero and finite $\tau$ as $xy\to 0$.
This is not a standard short-distance singularity, which arises from operators at coincident points, because it occurs at nonzero $\tau$. Instead, we interpret this singularity as a manifestation of the UV/IR mixing of the theory.

To understand the UV/IR mixing better, let us look at the two-point function \eqref{phidot-2pt-A.5} for $x\rightarrow 0$ and nonzero $\tau,y$ with $\ell_x\rightarrow\infty$ and a finite but large $\ell_y$.  (Swapping $x$ and $y$ gives similar results.)  It is given by $-{1\over 2\pi^2} \sqrt{\mu\over\mu_0} {1\over\tau^2} \log \left({\ell_y\over 2\pi y}\right)$ in the third line of \eqref{dotphi2pt}.  We see that the singularity at $x=0$ is now regularized  by the IR cutoff $\ell_y$ in the $y$ direction. Intuitively, small $x$ corresponds to large $x$-direction momentum $p_x$. For nonzero $\tau$, such large $p_x$ contributions are suppressed unless the $y$-direction momentum $p_y$ is small. The IR cutoff $|p_y|>{2\pi\over \ell_y}$ prevents $p_y$ from being too small, and thus regularizes the answer.

We conclude that the correlation functions of $\partial_\tau\phi$ in the continuum limit are almost like those in a standard continuum field theory, but there are various unusual singularities signaling the UV/IR mixing.

\subsubsection{Thermodynamic limit}

In the thermodynamic limit \eqref{thermolimit}, we can approximate the sum by an integral with $k_i = 2\pi n_i /L_i$:
\ie
{1\over L_i} \sum_{n_i} \to {1\over 2\pi} \int_{-\pi}^{\pi} dk_i~.
\fe
The two point function becomes
\ie
&\langle \partial_\tau \phi_{\hat x,\hat y}(\tau) \partial_\tau \phi_{0,0}(0)\rangle
\\
& \rightarrow -{UK\over 2\pi^2}\sqrt{U\over K}  \int_{-\pi}^{\pi} dk_x  \int_{-\pi}^{\pi} dk_y~
\left| \sin\left({k_x\over 2}\right) \sin\left({k_y\over 2}\right)\right|
e^{ - | \sin({k_x\over 2})\sin({k_y\over 2}) | T}\cos(k_x\hat x)\cos(k_y\hat y)~.
\fe
The integral is evaluated in Appendix \ref{sec:app_thermo2pt}.
We summarize the correlation function in various limits below:
\ie\label{eq:2pt_thermo}
&\langle \partial_\tau \phi_{\hat x,\hat y}(\tau) \partial_\tau\phi_{0,0}(0)\rangle \to-{1\over 2\pi^2}\sqrt{U\over K}\dfrac{1}{\tau^2} g\left(\sqrt{UK}\tau,\hat x,\hat y \right)~,
\\
&g\left(\sqrt{UK}\tau,\hat x,\hat y \right)=
\begin{cases}
	\dfrac{UK\tau^2}{\hat x^2\hat y^2}\quad &\sqrt{UK}\tau\ll 1 \ ,\ \hat x\gg1 \ ,\ \hat y\gg 1
	\\
	\log\left(\dfrac{\sqrt{UK}\tau}{\hat x\hat y}\right)  \quad\ & \sqrt{UK}\tau\gg\hat x\hat y\gg1\ ,\ \hat x , \hat y\gg 1
	\\
	\log\left(\dfrac{4\sqrt{UK}\tau}{\hat y}\right)  \quad & \sqrt{UK}\tau\gg\hat y\gg 1 \ ,\ \hat x=0
	\\
	\log\left(\dfrac{4\sqrt{UK}\tau}{\hat x}\right)  \quad & \sqrt{UK}\tau\gg \hat x\gg 1 \ ,\ \hat y=0
	\\
	\log\left(16\sqrt{UK}\tau\right)  \quad & \sqrt{UK}\tau\gg 1 \ ,\ \hat x=0 \ ,\ \hat y=0
	\\
	\dfrac{4UK\tau^2}{\hat y^2}\quad &\sqrt{UK}\tau\ll 1 \ ,\ \hat x=0 \ ,\ \hat y\gg 1
	\\
	\dfrac{4UK\tau^2}{\hat x^2}\quad &\sqrt{UK}\tau\ll 1 \ ,\ \hat x\gg1 \ ,\ \hat y=0
	\\
	16UK\tau^2\quad &\sqrt{UK}\tau\ll 1 \ ,\ \hat x=0 \ ,\ \hat y=0
\end{cases}
\fe

Let us rewrite the correlation functions in the thermodynamic limit using continuum variables and compare them with those in the continuum limit. In terms of continuum variables, the correlation functions with $xy\neq0$ in \eqref{eq:2pt_thermo} and \eqref{dotphi2pt}  agree with each other. The fact that the two computations can be continued to each other is because the plane wave spectrum is continuous at energies of order $1/L$. However, the correlation functions on the $xy=0$ locus in  \eqref{eq:2pt_thermo} and \eqref{dotphi2pt} do not agree with each other. In contrast to the continuum expression \eqref{dotphi2pt}, the singularities at $xy=0$ is regularized by the lattice space $a$ in \eqref{eq:2pt_thermo}.  This is another manifestation of the UV/IR mixing.

\subsection{$\langle e^{i\phi} e^{-i\phi} \rangle$}\label{sec:monopole}
Consider the two-point function of the monopole operator $e^{i\phi}$.\footnote{We refer to $e^{i\phi}$ as a monopole operator, and to $e^{i\phi_{0,0}(\tau)-i\phi_{\hat x,0}(\tau)}$, a normal-ordered bi-local operator made of $e^{i\phi}$, as a dipole operator. They should not be confused with magnetic monopoles or magnetic dipoles.} Since the monopole operator is charged under the subsystem symmetry, its two-point function is nonzero only when  the two operators are at the same spatial coordinates:
\ie\label{mon-2pt}
&\langle e^{i\phi_{0,0}(\tau)} e^{-i\phi_{0,0}(0)} \rangle = e^{-F(0,0,\tau)}~,
\\
&F(0,0,\tau) = { U \over2 L_xL_y} \left(L_x +L_y -1\right)|\tau| + {U\over 2L_xL_y }
 \sum_{-{L_x\over 2} <n_x\le {L_x\over2} \atop n_x\neq 0} \sum_{-{L_y\over 2} <n_y\le {L_y\over2} \atop n_y\neq 0}   {1- e^{-\omega_{n_x,n_y}|\tau|}\over \omega_{n_x,n_y}}~.
\fe

\subsubsection{Continuum limit}

In the continuum limit \eqref{contlimit}, only the first term  $F^{(0)}(0,0,\tau)$ contributes, and we have
\ie\label{monopole2ptlimit2}
\langle e^{i\phi_{0,0} (\tau) } e^{-i \phi_{0,0} (0)}\rangle \sim \exp\left[-{ 1\over2 \mu_0 \ell_x\ell_y} \left ({ \ell_x\over a}+{\ell_y\over a}-1\right)|\tau|\right]~,
\fe
which is dominated by exchanging the lowest energy charged state created by $e^{i\phi}$.   This agrees with equation (A.11) of \cite{paper1}.
Since the two-point function of $e^{i\phi}$ is nonzero only if the two operators are located at the same point in space, unlike the analysis in Section \ref{sec:phidot}, we cannot regularize this correlator by separating the operators in $x$ or $y$.
Rather, the short distance singularity must be regularized by the lattice spacing $a$.

The two-point function vanishes in the limit $a\to0$, and hence the  operator $e^{i\phi}$ does not act in the continuum limit.  It is a redundant operator.  This is consistent with our discussion of the low-energy theory in Section \ref{sec:cont}.
Another consequence of this fact is that the $U(1)$ momentum subsystem symmetry, under which $e^{i\phi}$ is charged, is  not spontaneously broken.  Using the self-duality of the system, the same conclusion applies to the $U(1)$ winding subsystem symmetry.

\subsubsection{Thermodynamic limit}

In the thermodynamic limit \eqref{thermolimit}, we can replace ${1\over L_i} \sum_{n_i}$ by ${1\over 2\pi} \int_{-\pi}^{\pi} dk_i$:
\ie
F(0,0,\tau) \to {1\over 8(2\pi)^2}  \sqrt{ U\over K}
\int_{-\pi}^{\pi} dk_x\int_{-\pi}^{\pi} dk_y~{1\over |\sin ({k_x\over2})\sin ({k_y\over2}) |}
\left(
1- e^{- |\sin( {k_x\over2} )\sin( {k_y\over2} )|T}
\right)~.
\fe
The integral can be expressed in terms of the generalized hypergeometric functions.\footnote{The integral is
\ie
&\int_{-\pi}^{\pi} dk_x\int_{-\pi}^{\pi} dk_y~{1\over |\sin ({k_x\over2})\sin ({k_y\over2}) |}
\left( 1- e^{- |\sin( {k_x\over2} )\sin( {k_y\over2} )|T} \right)
\\
&~~~= 4\pi^2T \,_2F_3\left(\frac{1}{2},\frac{1}{2};1,1,\frac{3}{2};\frac{T^2}{4}\right)-8T^2 \,_3F_4\left(1,1,1;\frac{3}{2},\frac{3}{2},\frac{3}{2},2;\frac{T^2}{4}\right)~.
\fe
}

Let us consider the limit $T\gg 1$. In this limit, the integral becomes $8 (\log T)^2  + {\cal O}(\log T)$.
Therefore,
\ie\label{logsquare}
\langle e^{i\phi_{0,0} (\tau) } e^{-i \phi_{0,0} (0)}\rangle \sim \exp \left[-  {1\over (2\pi)^2} \sqrt{U\over K} \left(\log (4\sqrt{UK}|\tau|)\right)^2  \right]~.
\fe
This agrees with equation (58) of \cite{PhysRevB.66.054526}.  This implies that the two-point function of the momentum operator decays faster than any power law.\footnote{Similar functional forms appeared recently in \cite{Shackleton:2021fdh}.  This scaling violation is also reminiscent of \cite{Metlitski:2020cqy}.}

In terms of the continuum variables \eqref{contvariable}, the two-point function
\ie
\langle e^{i\phi_{0,0} (\tau) } e^{-i \phi_{0,0} (0)}\rangle \sim \exp \left[-  {1\over (2\pi)^2} \sqrt{\mu\over \mu_0} \left(\log \left({4|\tau|\over\sqrt{\mu\mu_0}a^2}\right)\right)^2  \right]
\fe
depends on the lattice spacing $a$. Note that this dependence cannot be absorbed into the wavefunction renormalization of $e^{i\phi}$.  Furthermore, this correlation function does not respect the scaling symmetry \eqref{scalingxy} of the continuum action.  Indeed, it vanishes as $a\to 0$.

We conclude that the two-point function of the monopole operators \eqref{mon-2pt} vanishes in the $a\to0$ limit regardless of how we take $L\to \infty$.  This is consistent with the fact that the monopole operator does not act in the low-energy theory.

\subsection{$\langle e^{i\phi_{0,0}(\tau)-i\phi_{\hat x,0}(\tau)} e^{-i\phi_{0,\hat y}(0)+i\phi_{\hat x,\hat y}(0)} \rangle$}\label{sec:dipole}

Next, we consider the two-point function of the dipole operator  separated in, say, $x$ direction:
\ie\label{dip-2pt}
&\langle e^{i\phi_{0,0}(\tau)-i\phi_{\hat x,0}(\tau)} e^{-i\phi_{0,\hat y}(0)+i\phi_{\hat x,\hat y}(0)} \rangle = e^{- 2F(0,\hat y,\tau) + 2F(\hat x,\hat y,\tau)}\\
&=\exp\left[
- {U\over L_y}|\tau|-
{\sqrt{U/K} \over 2L_xL_y }\sum_{-{L_x\over2} < n_x\le {L_x\over2} \atop n_y\neq0} \sum_{-{L_y\over2} < n_y\le {L_y\over2} \atop n_y\neq0}
{1- e^{-\left| \sin {n_x \pi\over L_x} \sin {n_y \pi\over L_y}\right  |T}  \cos\left( {2\pi n_y \hat y\over L_y}\right)
\over| \sin {n_x \pi\over L_x} \sin {n_y \pi\over L_y} |}
\sin^2 \left({\pi n_x\hat x\over L_x}\right)
\right]\,.
\fe
The notation $e^{i\phi_{0,0}(\tau)-i\phi_{\hat x,0}(\tau)}$ for the dipole means certain normal-ordering  of the two monopoles.
The dipole two-point function \eqref{dip-2pt} is related to the monopole four-point function \eqref{mon-4pt-1} (with $\tau_1 = 0$ and $\tau_2 = \tau_3 = -\tau$) by removing the self-contraction terms within the dipole operator.

\subsubsection{Continuum limit}

In the continuum limit \eqref{contlimit}, only the linear term in $|\tau|$ in the exponent of \eqref{dip-2pt} contributes, and we have
\ie\label{eq:dipole2pt_conti}
&\langle e^{i\phi_{0,0}(\tau)-i\phi_{\hat x,0}(\tau)} e^{-i\phi_{0,\hat y}(0)+i\phi_{\hat x,\hat y}(0)} \rangle
\sim \exp\left(
- {1\over \mu_0 \ell_y a}|\tau|
\right)\,.
\fe
The two-point function in this limit is controlled by the lowest energy charged state.
 This agrees with equation (A.16) of \cite{paper1}.
The two-point function vanishes in the limit $a\to0$, and hence the dipole operator $e^{i\phi_{0,0}-i\phi_{\hat x,0}}$ does not act in the continuum.  It is a redundant operator.

\subsubsection{Thermodynamic limit}\label{sec:dipolethermo}

The thermodynamic limit \eqref{thermolimit}  probes the charged states. In this limit, we can approximate the sums by integrals ${1\over L_i} \sum_{n_i}$ by ${1\over 2\pi} \int_{-\pi}^{\pi} dk_i$:
\ie
&\langle e^{i\phi_{0,0}(\tau)-i\phi_{\hat x,0}(\tau)} e^{-i\phi_{0,\hat y}(0)+i\phi_{\hat x,\hat y}(0)} \rangle
\\
&=\exp\left[-
{1 \over 2(2\pi)^2 } \sqrt{\frac{U}{K}}\int_{-\pi}^{\pi} dk_x \int_{-\pi}^{\pi} dk_y~
{1- e^{-| \sin {k_x\over 2} \sin {k_y\over 2} |T}  \cos\left( {k_y \hat y}\right)
	\over| \sin {k_x\over 2} \sin {k_y\over 2} |}
\sin^2 \left({k_x\hat x\over 2}\right)\right]~.
\fe
The integral can be expressed in terms of harmonic numbers and regularized generalized hypergeometric functions.\footnote{
The integral is
\ie
&\int_{-\pi}^{\pi} dk_x \int_{-\pi}^{\pi} dk_y~
{1- e^{-| \sin {k_x\over 2} \sin {k_y\over 2} |T}  \cos\left( {k_y \hat y}\right)
	\over| \sin {k_x\over 2} \sin {k_y\over 2} |}
\sin^2 \left({k_x\hat x\over 2}\right)
\\
&~~~=8(H_{\hat x} - 2 H_{2\hat x})(H_{\hat y} - 2 H_{2\hat y})
\\
&\quad + \frac{\pi^{\frac32}}{2}T\cos(\pi \hat y) \left[ 2\pi \,\,_2\tilde F_3\left(\frac12,\frac12;\frac32,1-\hat y,1+\hat y;\frac{T^2}{4}\right) - T\, \,_3\tilde F_4\left(1,1,1;\frac32,2,\frac32-\hat y,\frac32+\hat y;\frac{T^2}{4}\right) \right]
\\
&\quad - \frac{\pi^2}{4}T\cos(\pi \hat x)\cos(\pi \hat y) \left[ 4\sqrt\pi\,\,_4\tilde F_5\left(\frac12,\frac12,1,1;\frac32,1-\hat x,1+\hat x,1-\hat y,1+\hat y;\frac{T^2}{4}\right) \right.
\\
&\qquad \qquad \qquad \qquad \qquad \qquad  - \left.T\,\,_4\tilde F_5\left(1,1,1,\frac32;2,\frac32-\hat x,\frac32+\hat x,\frac32-\hat y,\frac32+\hat y;\frac{T^2}{4}\right) \right]~,
\fe
where $H_n = \sum_{k=1}^n \frac1k$ is the $n$th harmonic number.
}

Next, we take $\hat y,T\gg 1$ with fixed $\hat y/T$ and  obtain
\ie\label{eq:dipole2pt}
&\langle e^{i\phi_{0,0}(\tau)-i\phi_{\hat x,0}(\tau)} e^{-i\phi_{0,\hat y}(0)+i\phi_{\hat x,\hat y}(0)} \rangle
\\
&=\exp\left[-
{1 \over (2\pi)^2 } \sqrt{\frac{U}{K}}\int_{-\pi}^{\pi} dk_x ~
\frac{\sin^2 \left({k_x\hat x\over 2}\right)}{|\sin {k_x\over 2}|}
\log\left(4UK\sin^2\left(\frac{k_x}{2}\right)\tau^2+\hat y^2\right)
\right]
\,.
\fe
In this limit, the correlation function develops a scale symmetry of $y$:
$\tau \rightarrow \lambda\tau$, $\hat y\rightarrow \lambda\hat y$.  Under this transformation, the dipole operator $e^{i\phi_{0,0}-i\phi_{\hat x,0}}$  transforms with a scaling dimension
\ie\label{eq:exponent}
\Delta_{\hat x} = {1  \over (2\pi)^2 }\sqrt{\frac{U}{K}} \int_{-\pi}^{\pi} dk_x~
{ \sin^2\left( {k_x \hat x\over 2} \right)\over |\sin{k_x\over2}| }=\frac{1}{\pi^2}\sqrt{\frac{U}{K}}\sum_{m_x=1}^{\hat x} {1\over 2m_x-1}~.
\fe

For $\hat y=0$ and $T\gg 1$, we have
\ie\label{dipolepowerlaw}
&\langle e^{i\phi_{0,0}(\tau)-i\phi_{\hat x,0}(\tau)} e^{-i\phi_{0,0}(0)+i\phi_{\hat x,0}(0)} \rangle
\sim\exp\left[
- {\sqrt{U/K} \over (2\pi)^2}
\log\left (  T^2\right)
\int_{-\pi}^\pi dk_x
{ \sin^2\left( {k_x \hat x\over 2} \right)\over |\sin {k_x\over2}| }
\right]\sim {1\over |\tau|^{2\Delta_{\hat x}}}~.
\fe
This agrees with Equation  (121) of \cite{PhysRevB.66.054526}.
For large $\hat x$, $\Delta_{\hat x}$ grows as
\ie
\Delta_{\hat x}\sim\frac{1}{2\pi^2}\sqrt{\frac{U}{K}}\log\hat x~.
\fe
We see that this correlation function does not satisfy the scale symmetry of $x$, which acts on the coordinates as $\tau\rightarrow \lambda\tau$, $\hat x\rightarrow\lambda\hat x$.

To understand the scaling symmetry in $y$ better, let us generalize \eqref{contvariable} to have different lattice spacings $a_x,a_y$ in the two spatial directions
\ie
U = {1\over \mu_0 a_x a_y}\,,~~~K = {1\over \mu a_x a_y} \,,~~~
L_i  ={ \ell_i\over a_i}\,,~~~~T=  {4\over \sqrt{\mu\mu_0}} {|\tau|\over a_x a_y}\,,~~~~\hat x^i = {x^i\over a_i}\,.
\fe
We can consider taking either the continuum limit $a_i\rightarrow 0$ or the thermodynamic limit $\ell_i\rightarrow\infty$. For the dipole two-point functions, the continuum limit $a_y\rightarrow 0$ in the $y$ direction and the thermodynamic limit $\ell_x\rightarrow \infty$ in the $x$ direction commute with all the other limits. Hence we can first take the two limits, in which, the action becomes
\ie\label{ycontaction}
S=\sum_{\hat x}\int dt dy\left[ \frac{\mu_0 a_x}{2}(\partial_\tau \phi)^2+\frac{1}{2\mu a_x}(\Delta_x \partial_y\phi)^2\right]~.
\fe

This action describes a system of coupled wires, where each wire supports a 1+1 dimensional theory of a compact boson. It has a separate $2\pi$ identification for the bosons on each wire $\phi(\tau,\hat x,y)\sim \phi(\tau,\hat x,y)+2\pi n_x(\hat x)$ as well as the identification $\phi(\tau,\hat x,y)\sim \phi(\tau,\hat x,y)+2\pi n_y(y)$.  This theory has the scaling symmetry of $y$, but not in $x$.

As a check, the $a_y\to 0$ limit of the dipole two-point function can be computed directly using the continuum action \eqref{ycontaction}.

Next, we can try to take the continuum limit in the $x$ direction, i.e., turn the discrete wires in \eqref{ycontaction} into a continuous plane.  Then, we find that the  $a_x\rightarrow 0$ and the $\ell_y\rightarrow\infty$ limits do not commute.
In the $a_x\rightarrow 0$ limit, the result agrees with the continuum limit \eqref{eq:dipole2pt_conti}, while in the $\ell_y\rightarrow \infty$ limit, the result agrees with the thermodynamic limit \eqref{eq:dipole2pt} written in terms of the continuum variables.

\subsection{$\langle e^{i\phi} e^{-i\phi} e^{i\phi} e^{-i\phi} \rangle$}
The subsystem symmetry allows two nontrivial four-point functions of the monopole operator $e^{i\phi}$. One of them is
\ie\label{mon-4pt-1}
&\langle e^{i\phi_{0,0}(0)} e^{-i\phi_{\hat x,0}(\tau_1)} e^{-i\phi_{0,\hat y}(\tau_2)} e^{i\phi_{\hat x,\hat y}(\tau_3)} \rangle
\\
&~~~= \exp[- F(\hat x,0,\tau_1) - F(0,\hat y,\tau_2) + F(\hat x,\hat y,\tau_3)
\\
&  \quad \qquad + F(\hat x,\hat y,\tau_1-\tau_2) - F(0,\hat y,\tau_1-\tau_3) - F(\hat x,0,\tau_2-\tau_3) ]~,
\fe
and the other is
\ie
&\langle e^{i\phi_{0,0}(0)} e^{-i\phi_{\hat x,\hat y}(\tau_1)} e^{-i\phi_{0,0}(\tau_2)} e^{i\phi_{\hat x,\hat y}(\tau_3)} \rangle
\\
&~~~= \exp[- F(\hat x,\hat y,\tau_1) - F(0,0,\tau_2) + F(\hat x,\hat y,\tau_3)
\\
&  \quad \qquad + F(\hat x,\hat y,\tau_1-\tau_2) - F(0,0,\tau_1-\tau_3) - F(\hat x,\hat y,\tau_2-\tau_3) ]~.
\fe

Let us focus on a special case of the first four-point function where $\tau_1 = \tau_2 = \tau_3 = 0$:
\ie
&\langle e^{i\phi_{0,0}(0)} e^{-i\phi_{\hat x,0}(0)} e^{-i\phi_{0,\hat y}(0)} e^{i\phi_{\hat x,\hat y}(0)} \rangle = e^{- 2F(\hat x,0,0) - 2F(0,\hat y,0) + 2F(\hat x,\hat y,0) }\\
&=\exp\left[
-{1\over L_xL_y} \sqrt{U\over K}
\sum_{-{L_x\over2} < n_x \le {L_x\over 2} \atop n_x\neq0 }
{ \sin^2\left({\pi n_x \hat x\over L_x}\right)\over |\sin\left({\pi n_x\over L_x}\right)|}
\sum_{-{L_y\over2} < n_y \le {L_y\over 2} \atop n_y\neq0 }
{ \sin^2\left({\pi n_y \hat y\over L_y}\right)\over |\sin\left({\pi n_y\over L_y}\right)|}
\right]\,.
\fe
In the thermodynamic limit $L_i\to\infty$, we can approximate the sums by integrals and rewrite it as
\ie
&\langle e^{i\phi_{0,0}(0)} e^{-i\phi_{\hat x,0}(0)} e^{-i\phi_{0,\hat y}(0)} e^{i\phi_{\hat x,\hat y}(0)} \rangle
= \exp\left[
-{1\over \pi^2} \sqrt{U\over K}
\sum_{m_x=1}^{\hat x} {1\over 2m_x-1}
\sum_{m_y=1}^{\hat y} {1\over 2m_y-1}
\right]\,.
\fe
If we further take the large $\hat x,\hat y\gg 1$ limit, we obtain
\ie
&\langle e^{i\phi_{0,0}(0)} e^{-i\phi_{\hat x,0}(0)} e^{-i\phi_{0,\hat y}(0)} e^{i\phi_{\hat x,\hat y}(0)} \rangle
\sim \exp\left[ - {1\over \pi^2} \sqrt{U\over K} \log \hat x\log \hat y
\right]\,,
\fe
which agrees with equation (66) of \cite{PhysRevB.66.054526}.
In terms of the continuum variables $x ^i=\hat x^i a$, the four-point function depends on the lattice spacing $a$ and this dependence cannot be rescaled into the wavefunction renormalization of the operator.

Again, this signals the UV/IR mixing and the loss of scaling symmetry when we include these charged operators as discussed in Section  \ref{sec:spectrum}.

\subsection{Lessons from the correlation functions}

Let us summarize our findings.

The continuum field theory \eqref{eq:action_continuum} captures correctly the physics of the lowest-lying states, the plane wave states.  Correspondingly, the correlation functions of operators of the form $\partial_\tau \phi$ or $\partial_x \partial_y\phi$ are almost standard.  The only novelty in these correlation functions is their UV/IR mixing.  In the infinite volume limit $\ell_i\to \infty$, the correlation functions are singular on certain spatial submanifolds.  Even though these singularites appear to be UV singularities, they are regularized in the finite volume theory.

The correlation functions of exponential operators are more subtle.  They are associated with states that are present in the underlying lattice theory, but are not included in the low-energy theory of Section \ref{sec:cont}.  Since these states are the lowest energy states carrying charges under conserved global symmetries, they are still interesting \cite{paper1}.  As we explained in Section \ref{sec:cont}, these states can be interpreted as defects in the low-energy theory.  Their correlation functions exhibit more dramatic UV/IR mixing than the operators in the low-energy theory.

The fact that the exponential operators like the monopoles $e^{i\phi_{\hat x,\hat y}(\tau)}$ and $e^{i\phi^{xy}_{\hat x,\hat y}(\tau)}$ and the dipoles $e^{i[\phi_{\hat x,\hat y}(\tau)-\phi_{\hat x',\hat y}(\tau)]}$, do not act in the low-energy continuum theory  has important implications.  We can explicitly break the momentum and winding subsystem symmetries by deforming the UV theory by such operators.  If the coefficient of this deformation is sufficiently small, there is no effect on the low-energy theory.  These operators are infinitely irrelevant.  Consequently, the momentum and winding subsystem global symmetries are not spontaneously broken in the large volume limit \cite{PhysRevB.72.045137,You:2019cvs,paper1}.  Furthermore, for sufficiently small coefficient, the low-energy theory is robust.  It has the same spectrum of plane waves and the same correlation functions.  This is the case despite the fact that the global symmetries are explicitly violated at short distances \cite{paper1}.  This point will be discussed further in Section \ref{sec:dipole-deform}.

\section{Deformations and the 't Hooft anomaly matching}\label{sec:dipole-deform}

In this section, we discuss further the deformation of the modified Villain model \eqref{eq:S-Villain-lattice} by  the charged operators. These deformations of the original XY-plaquette model in the Hamiltonian formalism were discussed in \cite{PhysRevB.66.054526} and in the continuum $\phi$-theory in \cite{paper1}.

As we reviewed in Section \ref{sec:correlation}, the perturbations by the momentum operator $e^{i\phi}$ or by the winding operator $e^{i\phi^{xy}}$ are infinitely irrelevant in the continuum theory.
As a result, the low-energy theory is robust under small deformations of the UV theory by these operators.

Next, let us consider deformations of the modified Villain model \eqref{eq:S-Villain-lattice} by the dipole operators
\ie\label{dipoledef}
-\varepsilon\sum_\text{links}\cos(\Delta_i\phi)~.
\fe
 In the continuum limit,  the dipole operators do not act.
 In the thermodynamic limit, on the other hand,
the two-point function \eqref{dipolepowerlaw} exhibits  a power-law decay on the $\tau y$-plane.
However,
the exponent \eqref{eq:exponent} of the power-law decay does not directly determine the relevance or the irrelevance of the dipole operator.  One way to see that is that the dipole operator has simple scaling in one spatial direction in \eqref{scalingxy}, but not in the other spatial direction in \eqref{scalingxy}. See Section \ref{sec:dipolethermo}.

 Below we will study the deformed theory by taking a different approach based on the global symmetry and the 't Hooft anomaly, and show that the low-energy phase has to be gapless. We propose some candidate continuum theories for the gapless phases.

The dipole deformation breaks the $U(1)$ momentum subsystem symmetry to an ordinary (zero-form) $U(1)$ momentum  symmetry, which  shifts $\phi$ by  a constant, $\phi\rightarrow\phi+\alpha$.
On the other hand, it leaves the $U(1)$ winding subsystem symmetry intact.

Before the deformation, the $U(1)$ momentum and the $U(1)$ winding subsystem symmetries have a mixed 't Hooft anomaly \cite{Gorantla:2021svj}. (See Section \ref{sec:mV} and Appendix \ref{sec:anomaly_lattice} for a review.)
After the deformation, the unbroken symmetries are the (zero-form) $U(1)$ momentum symmetry and the $U(1)$ winding subsystem symmetry.  This global symmetry still has a residual 't Hooft anomaly.

This residual anomaly can be seen by coupling the momentum and winding symmetries to their classical background gauge fields.  This coupling and the anomaly are discussed on the lattice in Appendix \ref{sec:anomaly_lattice}.  In the continuum, the background gauge fields are $A_\mu$ with $\mu=\tau,x,y$ for the momentum symmetry and $(\tilde A^{xy}_\tau, \tilde A)$  for the winding subsystem symmetry.  Their gauge symmetries are $A_{\mu}\rightarrow A_\mu+\d_\mu \alpha$ and $(\tilde A^{xy}_{\tau},\tilde A)\rightarrow (A^{xy}_{\tau}+\d_\tau\tilde\alpha^{xy},\tilde A+\d_x\d_y\tilde\alpha^{xy})$, respectively. Here $\alpha$ and $\tilde\alpha^{xy}$ are the background gauge parameters.  In the continuum limit, the variation of the action under background gauge transformation is
\ie\label{XYplaq-deform-anomaly-lattice-continuum}
\frac{i}{2\pi}&\int d\tau dx dy\, \alpha(\d_\tau \tilde A - \d_x \d_y \tilde A_\tau^{xy} )~.
\fe
It signals an 't Hooft anomaly because it cannot be canceled by any 2+1d local counterterms.

The mixed 't Hooft anomaly  implies that the modified Villain lattice model has to remain gapless after deformation by the momentum dipole operators.

Let us comment on the implications for the 2+1d XY-plaquette model \eqref{eq:H_XY_plaq}.
The  XY-plaquette model with  $\mathcal K\gg \mathcal U$ is in the same phase as its modified Villain version with ${\cal K}\approx K,{\cal U}\approx U$.
When we turn on the deformation with nonzero $\varepsilon$, the modified Villain model remains gapless.  Hence, the same is true for the XY-plaquette model with $\mathcal K\gg \mathcal U$.

When the deformation is small $\varepsilon \ll 1$, one candidate  continuum field theory for this gapless phase is  \eqref{eq:action_continuum}.
It describes   the  low-lying plane wave states with energy of order $1/L^2$.  On the other hand,
the charged states with energy of order $1/L$,  are not robust under the dipole deformation.
This was demonstrated in \cite{PhysRevB.66.054526} by computing the second-order perturbation of the two-point function of $e^{i\phi}$.

When the deformation is large $\varepsilon\gg 1$, the theory appears to flow at long distances to the 2+1d compact boson  field theory, with the Euclidean action
\ie\label{2dcontt}
S=\frac{f}{2}\int d\tau dx dy\, (\d_\mu\phi)^2~,~~~\phi\sim \phi+2\pi\,.
\fe
(In Section \ref{sec:3+1d}, we will discuss this theory from another perspective.) Intuitively, this is because the dipole deformation becomes the standard nearest-neighbor interaction of the 2+1d XY model.\footnote{There might or might not be a phase transition as $\varepsilon$ is tuned from zero to a large value.  Our point here is only that the model is gapless for any $\varepsilon$.}

To check this proposal for the gapless phase of the system, we have to check that the latter matches the residual 't Hooft anomaly \eqref{XYplaq-deform-anomaly-lattice-continuum} of the deformed microscopic lattice model.
We do it by coupling the compact boson theory to the same background gauge fields $A_\mu$ and $(\tilde A_\tau^{xy},\tilde A)$. The action after coupling is given by
\ie
S=\int d\tau dx dy\left[\frac{f}{2} (\d_\mu\phi-A_\mu)^2-\frac{i}{2\pi} \left(\tilde A_\tau^{xy}\d_x\d_y\phi+\tilde A \d_\tau\phi\right)\right]~.
\fe
It is shifted by \eqref{XYplaq-deform-anomaly-lattice-continuum} under the background gauge transformation. Hence, the 't Hooft anomaly of the deformed miscroscopic lattice model is matched by the compact boson theory.

We can further deform the microscopic lattice model by a winding dipole operators $\cos(\Delta_x\phi^{xy})$, $\cos(\Delta_y\phi^{xy})$. This breaks the $U(1)$ winding subsystem symmetry to a $U(1)$ winding zero-form symmetry.
However, there is no mixed 't Hooft anomaly between the two residual $U(1)$ zero-form symmetries, and therefore we cannot use it to constrain the low-energy spectrum.

After the completion of this work, \cite{Lake2021rg} proposed a renormalization group procedure for the thermodynamic limit of this model.

\section{3+1d theories}\label{sec:3+1d}

In this section, we will discuss some exotic 3+1d gapless theories.
Using the terminology of \cite{paper2,Gorantla:2021svj}, these include the $\phi$-theory (which is dual to the $\hat A$-theory) and the $\hat \phi$-theory (which is dual to the $A$-theory). We place all these lattice models on a Euclidean lattice with infinitely many sites in the Euclidean time direction, and $L=L_x=L_y=L_z$ sites in the spatial directions.

The discussion of these models will expose an important subtlety in what we mean by the low-energy limit and the continuum limit.  This subtlety is not present in our previous discussion of the 2+1d XY-plaquette model \eqref{eq:iH_XY_plaq} (or its modified Villain version \eqref{eq:iS-Villain-lattice}).  There, the low-energy limit is described by the continuum field theory \eqref{eq:i action_continuum}.  As we emphasized above, this description is valid for fixed values of the lattice coupling constants $\beta_0$ and $\beta$ \eqref{UKmumuzbeta}, provided we scale $a_\tau \sim a^2\to 0$.  In that limit, the continuum parameters $\mu_0= {\beta_0 a_\tau \over a^2}$ and $\mu={a_\tau \over \beta a^2}$ are fixed.  Equivalently, we focused on the low-energy states by scaling the parameters in the Hamiltonian $U$ and $K$, as $U,K \sim 1/a^2$, with fixed ratio $U/K$.

This will not be the case in the models in this section.  Here, there are different natural ways to scale the lattice parameters.  In the following subsection, we will demonstrate it in a well-known example.

\subsection{Review of the 2+1d XY model and its continuum limit}\label{sec:2+1dXY}

In order to demonstrate the issue we will face below, let us review briefly the situation in the 2+1d XY model.  (Compare with the 1+1d version of this system in Appendix \ref{onepone}.) It involves circle-valued degrees of freedom $\phi$ on a spatial lattice with Hamiltonian
\ie\label{3dXY}
H  = { \mathcal{U} \over2} \sum_{\text{site}} \pi^2 - {\mathcal{K} }\sum_{\text{links}}\cos(\Delta_i\phi)~.
\fe
This theory is gapless for large enough $\mathcal K/\mathcal U$ and is gapped for small $\mathcal K/\mathcal U$ with an interacting critical point between them.

A related lattice model is the modified Villain model \cite{Gorantla:2021svj} with the Euclidean lattice action
\ie\label{3dXYmo}
S=\frac{\beta}{2}\sum_{\mu\text{-link}}(\Delta_\mu \phi-2\pi n_\mu)^2+i\sum_{\nu\rho-\text{plaq}}\epsilon^{\mu\nu\rho}\tilde A_\mu(\Delta_\nu n_{\rho}-\Delta_\rho n_\nu)~,
\fe
where $\phi$ is a real-valued field on the sites, $n_\mu$ is an integer-valued gauge field on the $\mu$-links, and $\tilde A_\mu$ is a circle-valued gauge field on the dual $\mu$-links, which acts as Lagrange multiplier enforcing the flatness of $n_\mu$.  The model based on this action is gapless for all values of $\beta$ \cite{Gorantla:2021svj}.  And for large $\beta$, it is closely related to the model based on \eqref{3dXY} in its gapless phase.  Furthermore, that behavior is captured by the continuum model based on
\ie\label{3dXYcon}
S=\frac{f}{2}\int d\tau dx dy\, (\d_\mu\phi)^2~,~~~\phi\sim \phi+2\pi\,.
\fe

As in the discussion in the introduction, we relate the parameters in all these models.  The long-time correlation functions of the Euclidean model based on \eqref{3dXYmo} can be described by the action
\ie\label{quadac}
&S=\int d\tau \left[{1\over 2U}\sum_{\rm sites}(\partial_\tau \phi)^2 + {K\over 2} \sum_{\rm links} (\Delta_i \phi)^2\right]~, \\
&U={1\over \beta a_\tau}~,\quad K={\beta \over a_\tau}~,
\fe
where $a_\tau$ is the lattice spacing in the time direction and we suppressed the integer-valued gauge fields.  Comparing with \eqref{3dXY}, we identify $U\approx {\mathcal U}$ and $K \approx {\mathcal K}$ for large ${K/ U}=\beta^2$.  Next, we consider correlation functions at large spatial separations.  Here we recover the effective action \eqref{3dXYcon}.  Since the model is Lorentz invariant, we set $a_\tau=a$ and conclude that the parameters are related by
\ie\label{UKmumuzbetas}
f=K= {1\over Ua^2}={\beta\over a}~.
\fe
Note that Lorentz invariance leads to $UK={1/ a^2}$ and therefore it diverges in the continuum limit.

Repeating the analysis of the spectrum in Section \ref{sec:spectrum}  for this theory, we find
\ie\label{states3d}
&{\rm plane\ waves}\qquad \qquad &&{\sqrt{UK}\over L} ={1\over \ell} ~,\\
&{\rm momentum\ states}\qquad &&{U\over L^2}={1\over \beta a L^2}={1\over f \ell^2}~,\\
&{\rm winding\ states} \qquad \quad &&K={\beta\over a}=f~,
\fe
with, as above, $\ell =L a$.

What do we mean by the low-energy limit?

One possible answer involves taking $L\to \infty$, while holding $U$ and $K$ fixed, and focusing on the lowest lying states.  Here, we find only the momentum states with energy of order $U/ L^2$.  For large $L$, these states approach zero energy and for infinite $L$, they can mix.  The Hilbert space splits to superselection sectors labeled by $\langle \phi\rangle$, rather than states with fixed momentum.  This is the statement that the momentum global symmetry is spontaneously broken.  Therefore, focusing only on the momentum states and taking $L\to \infty$, we end up with decoupled Hilbert spaces with a single state in each of them.

A more refined limit takes into account not only the momentum states with energy of order $1/ L^2$, but also the plane waves with energy of order $1/ L$.  We scale $L\to \infty$, with $a\to 0$, holding $\ell=aL$ fixed.  Keeping $U/ K$ fixed, we simply scale our units of energy.  This means that $U \sim K\sim{1/ a}\to \infty$.  Equivalently, we can think of this limit is $a\to 0$ holding $\beta=aK={1/( aU)}$ fixed.  Therefore, below, we will refer to this limit as the ``fixed lattice couplings limit.''  In this limit, the physical length of the system $\ell$ is finite and the plane waves have finite energy of order $1/ \ell$.  The momentum states still go to zero energy, while the winding states scale to infinite energy.  In this limit, the boson becomes effectively non-compact.

Another interesting limit leads to the continuum theory \eqref{3dXYcon} with finite $f$.  Here, we again take $L\to \infty$ and  $a \to 0$ with fixed $\ell=aL$, but we also scale $\beta=f a \to 0$, such that the continuum coupling $f$ in \eqref{UKmumuzbetas} is held fixed.  Therefore, below, we will refer to this limit as the ``fixed continuum coupling limit.''  In this limit, all the states in \eqref{states3d} have finite energy. In terms of \eqref{quadac}, this limit corresponds to taking $K/U$ to zero.\footnote{This is far from the region where the original model \eqref{3dXY} is similar to the other models.  In terms of that model, this limit is the low-energy limit combined with a limit toward the critical point of the model.}

\subsection{3+1d $\phi$-theory/$\hat A$-theory}\label{sec:3+1dXYplaq}

The 2+1d XY-plaquette model of Section \ref{sec:2+1d-XYplaq} can be extended naturally to the 3+1d XY-plaquette model. Its modified Villain version is described by the action \cite{Gorantla:2021svj}
\ie\label{3+1d-phi-modVill-action}
&  \frac{\beta_0}{2} \sum_{\tau\text{-link}} (\Delta_\tau \phi - 2\pi n_\tau)^2 + \frac{\beta}{2} \sum_{i<j}\sum_{ij\text{-plaq}} (\Delta_i \Delta_j \phi - 2\pi n_{ij})^2
\\
&\quad + i \sum_{i<j}\sum_{\tau ij\text{-cube}} \hat A^{ij} (\Delta_\tau n_{ij} - \Delta_i \Delta_j n_\tau) - \frac i2 \sum_{xyz\text{-cube}}~\sum_{i\ne j\ne k} \hat A_\tau^{[ij]k} (\Delta_i n_{jk} - \Delta_j n_{ik})~,
\fe
where $\phi$ is a real scalar field, $(n_\tau,n_{ij})$ are integer-valued fields, and $(\hat A_\tau^{[ij]k},\hat A^{ij})$ are Lagrange multipliers that impose the flatness constraint of $(n_\tau,n_{ij})$.  A duality transformation makes $\hat A$ the dynamical field and turns $\phi$ into a Lagrange multiplier \cite{Gorantla:2021svj}.  We refer to the corresponding dual continuum theory as the $\hat A$-theory.

The continuum field theory corresponding to this lattice model is
\ie\label{3dphitheory}
\int d\tau dx dy dz ~ \left[ \frac{\mu_0}{2} (\partial_\tau \phi)^2 + \frac{1}{2\mu} \sum_{i<j} (\partial_i \partial_j \phi)^2 \right]~,
\fe
where $\phi$ is a compact scalar field. $\mu_0$ and $\mu$ has mass dimension +2 and 0, respectively.
This continuum field theory has been discussed in  \cite{Slagle:2017wrc,You:2018zhj,Radicevic:2019vyb,Gromov:2020yoc,paper2}.
It is dual to a tensor gauge theory of $\hat A$ of \cite{paper2} (see also \cite{Slagle:2017wrc,Radicevic:2019vyb}).

Repeating the analysis leading to \eqref{UKmumuzbeta}, we can relate the parameters in \eqref{3+1d-phi-modVill-action} and \eqref{3dphitheory}:
\ie
\beta_0 = \frac{\mu_0 a^3}{a_\tau}={\mu_0a\ell} ~, \qquad \beta = \frac{a_\tau}{\mu a}=\frac{a}{\mu\ell }~.
\fe
Here we set $a_\tau = a^2/\ell$ with $\ell=La$ so that, as we will see below, the energies of the plane waves do not depend on the lattice spacing $a$.

Following the same analysis as in Section \ref{sec:spectrum},  we find the spectrum of this lattice model:
\ie\label{phi-lat-energy}
&{\rm plane\ waves}\qquad \qquad &&E_\text{wave} \sim \frac{1}{a_\tau L^2}\sqrt{\frac{\beta}{\beta_0}}=\frac{1}{\sqrt{\mu\mu_0}\ell^2} ~,\\
&{\rm momentum\ states}\qquad &&E_\text{mom} \sim \frac{1}{a_\tau L^2\beta_0}=\frac{1}{\mu_0a\ell^2}~,\\
&{\rm winding\ states} \qquad \quad &&E_\text{wind} \sim \frac{\beta}{a_\tau}=\frac{1}{\mu a}~.
\fe
See Table \ref{tbl:lattice-energy} and \ref{tbl:cont2-energy} for a comparison with other theories.

Similar to the discussion in Section \ref{sec:2+1dXY}, we will consider two different low-energy limits: one with the lattice couplings fixed, and another with the continuum couplings fixed. We will discuss spontaneous symmetry breaking and robustness of these low-energy theories.

\begin{table}
\begin{center}
\begin{tabular}{|c|c|c|c|}
\hline
Theory & $E_\text{wave}$ & $E_\text{mom/elec}$ & $E_\text{wind/mag}$
\tabularnewline
\hline
3+1d $\phi$-theory & $\frac{1}{L^2}$ & $\frac{1}{L^2}$ & $1$
\tabularnewline
3+1d $\hat A$-theory & $\frac{1}{L^2}$ & $1$ & $\frac{1}{L^2}$
\tabularnewline
\hline
3+1d $\hat \phi$-theory & $\frac{1}{L}$ & $\frac{1}{L^2}$ & $1$
\tabularnewline
3+1d $A$-theory & $\frac{1}{L}$ & $1$ & $\frac{1}{L^2}$
\tabularnewline
\hline
\end{tabular}
\caption{
 The first two rows and the last two rows correspond to Sections \ref{sec:3+1dXYplaq} and \ref{sec:3+1dtensor}, respectively.
Note that the $L$ dependence of the energy of the modes in the XY-plaquette model differs from that in the 2+1d version of this model in Table \ref{tbl:lattice-energythr}.}\label{tbl:lattice-energy}
\end{center}
\end{table}

\begin{table}[t]
	\begin{center}
		\begin{tabular}{|c|c|c|c|c|}
			\hline
			Theory & $E_\text{wave}$ & $E_\text{mom/elec}$ & $E_\text{wind/mag}$ & $a_\tau$
			\tabularnewline
			\hline
			3+1d $\phi$-theory & $\frac{1}{\ell^2}$ & $\frac{1}{a \ell^2}$ & $\frac{1}{a}$ & $a^2$
			\tabularnewline
			3+1d $\hat A$-theory & $\frac{1}{\ell^2}$ & $\frac{1}{a}$ & $\frac{1}{a \ell^2}$ & $a^2$
			\tabularnewline
			\hline
			3+1d $\hat \phi$-theory & $\frac{1}{\ell}$ & $\frac{1}{a\ell^2}$ & $a$ & $a$
			\tabularnewline
			3+1d $A$-theory & $\frac{1}{\ell}$ & $a$ & $\frac{1}{a\ell^2}$ & $a$
			\tabularnewline
			\hline
		\end{tabular}
		\caption{The energies of the three kinds of states in various theories  in finite volume $\ell$ under the  limit $a \rightarrow 0$ with fixed continuum coupling constants. We ignore the dimensionful continuum coupling constants required to match the mass dimension. The first two rows and the last two rows correspond to the 3+1d exotic theories discussed in Sections \ref{sec:3+1dXYplaq} and \ref{sec:3+1dtensor}, respectively. The last column shows how to scale $a_\tau$ with $a$ in each theory.  Note that the $a$ and  $\ell$ dependence of the energy of the states in the $\phi$-theory differ from those in the 2+1d version of this model in Table \ref{tbl:cont2-energyth}.} \label{tbl:cont2-energy}
	\end{center}
\end{table}

\subsubsection{Fixed lattice couplings}

We first consider the fixed lattice coupling limit, where we keep the lattice parameters fixed, and take $L\rightarrow \infty$. In this limit, it follows from \eqref{phi-lat-energy} that plane waves and momentum states are equally light. Therefore, we cannot conclude whether the momentum symmetry is spontaneously broken by looking only at the spectrum. We need a more detailed analysis involving the correlation functions.

Following the steps used in Section \ref{sec:correlation}, the two-point function of the monopoles is
\ie
&\langle e^{i\phi(0,0,\tau)} e^{-i\phi(0,0,0)} \rangle = e^{-F(0,0,\tau)}~,
\\
&F(0,0,\tau) = \frac{1}{2\beta_0 a_\tau L_x L_y L_z} (L_x + L_y + L_z - 2) |\tau|
\\
&\qquad \qquad +  \frac{1}{2\beta_0 a_\tau L_x L_y L_z}  \sum_{-{L_x\over 2} <n_x\le {L_x\over2} \atop n_x\neq 0} \sum_{-{L_y\over 2} <n_y\le {L_y\over2} \atop n_y\neq 0} \sum_{-{L_z\over 2} <n_z\le {L_z\over2} \atop n_z\neq 0}  {1- e^{-\omega_{n_x,n_y,n_z}|\tau|}\over \omega_{n_x,n_y,n_z}}~,
\fe
where the dispersion relation is
\ie
\omega_{n_x,n_y,n_z} = \frac{4}{a_\tau} \sqrt{\frac{\beta}{\beta_0}} \sqrt{\sin^2\left({\pi n_x \over L_x}\right)\sin^2\left({\pi n_y \over L_y}\right) + \cdots}~,
\fe
where `$\cdots$' denotes similar terms in other directions.
The continuum limit corresponds to taking $L\rightarrow \infty$ while scaling  ${|\tau|}/{a_\tau} \sim L^2$ at the same time. In this limit, the first term of $F(0,0,\tau)$ remains finite, whereas the second term vanishes. If we further take the infinite volume limit $\ell \rightarrow \infty$, the first term also vanishes. Therefore, the two-point function approaches a non-zero constant at large $|\tau|$
\ie
\langle e^{i\phi(0,0,\tau)} e^{-i\phi(0,0,0)} \rangle \sim 1~,
\fe
which means the momentum symmetry is spontaneously broken.

We could also consider the limit $L \rightarrow \infty$ with fixed ${|\tau|}/{a_\tau}\sim 1$. Then, the first term of $F(0,0,\tau)$ vanishes, and the second term can be replaced by an integral:
\ie\label{finiteT}
F(0,0,\tau) \rightarrow -\frac{1}{(2\pi)^3\sqrt{\beta\beta_0}}  \int_0^\pi dk_x \int_0^\pi dk_y \int_0^\pi dk_z~  \frac{ e^{-T\sqrt{\sin^2\left({k_x \over 2}\right)\sin^2\left({k_y \over 2}\right) + \cdots}}}{\sqrt{\sin^2\left({k_x \over 2}\right)\sin^2\left({k_y \over 2}\right) + \cdots}}~,
\fe
where we defined $T\equiv \frac{4|\tau|}{a_\tau}\sqrt{\frac{\beta}{\beta_0}}$. When $T\gg 1$, we can approximate the integral by\footnote{In the large $T$ limit, the dominant contributions to the integral \eqref{finiteT} come from regions where the energy $\omega$ is small; in particular, $\omega \lesssim 1/T$. This occurs when at least two of the momenta are small.  The contribution of the region where all the three momenta are small, $k_x\sim k_y \sim k_z\sim 1/\sqrt{T}$, is of order ${1\over T^{3/2}} \times T={1\over T^{1/2}}$.  (Here, the first factor is the volume of this region in momentum space and the second is the integrand.)  This is to be compared with the contribution of the regions where only two of the momenta are small and the third is of order one, say, $k_x\sim k_y\sim 1/T$ and $k_z\sim 1$.  It is of order ${1\over T^2}\times T \sim {1\over T}$, which is smaller than the first contribution. Therefore, for large $T$, we can focus on the first region and approximate the integral \eqref{finiteT} by \eqref{largeT}.}
\ie\label{largeT}
F(0,0,\tau) &\rightarrow -\frac{1}{2\pi^3\sqrt{\beta\beta_0}}  \int_0^\infty dk_x \int_0^\infty dk_y \int_0^\infty dk_z~  \frac{ e^{-\frac{T}{4}\sqrt{k_x^2 k_y^2 + k_y^2 k_z^2 + k_z^2 k_x^2}}}{\sqrt{k_x^2 k_y^2 + k_y^2 k_z^2 + k_z^2 k_x^2}}
\\
& = - \frac{2\sqrt{2\pi} K(\frac12)^2}{2\pi^3\sqrt{\beta\beta_0}} \frac{1}{\sqrt{T}}~,
\fe
where $K(m)$ is the complete elliptic integral of the first kind.
If we further take the limit $a_\tau \rightarrow 0$, the propagator $F(0,0,\tau)$ vanishes, so, once again, we find that the two-point function approaches a non-zero constant at large $|\tau|$
\ie
\langle e^{i\phi(0,0,\tau)} e^{-i\phi(0,0,0)} \rangle \sim 1~,
\fe
which means the momentum symmetry is spontaneously broken.

The momentum symmetry is not robust because perturbing the theory by local operators charged under the momentum symmetry changes the low-energy spectrum. Operators charged under the winding symmetry are not local, so the winding symmetry is robust. See Table \ref{tbl:latticelargeL2} for a comparison with other theories.

\begin{table}[t]
	\begin{center}
		\begin{tabular}{|c|c|c|c|}
			\hline
			Theory & Light states & SSB & Robustness
			\tabularnewline
			\hline
			3+1d $\phi$-theory & Plane waves and momentum & Yes & Yes
			\tabularnewline
			3+1d $\hat A$-theory & Plane waves and magnetic & Yes & No
			\tabularnewline
			\hline
			3+1d $\hat \phi$-theory & Plane waves and momentum & Yes & Yes
			\tabularnewline
			3+1d $A$-theory & Plane waves and magnetic & Yes & No
			\tabularnewline
			\hline
		\end{tabular}
		\caption{The spectrum of the theories of Sections \ref{sec:3+1dXYplaq} and \ref{sec:3+1dtensor} in infinite volume with fixed lattice parameters. In all these theories, we consider whether the infinite volume theory exhibits spontaneous symmetry breaking of the momentum/magnetic symmetry. We also consider local perturbations violating the winding/magnetic symmetry, while imposing the momentum/electric symmetry.  In the table, we note whether the theory is robust under such deformations.
		}\label{tbl:latticelargeL2}
	\end{center}
\end{table}

\subsubsection{Fixed continuum couplings}

In the fixed continuum coupling limit \cite{paper2}, we take the limit $a\rightarrow 0$ while keeping the physical size of the system $\ell=aL$ and the continuum coupling constants $\mu_0,\mu$ fixed. In this limit, the theory is described by the continuum $\phi$-theory \eqref{3dphitheory}.

The continuum theory is not scale-invariant. Under the scale transformation $x^i \rightarrow \lambda x^i$ and $\tau \rightarrow \lambda^2 \tau$, the coupling constants transform as
\ie
\mu_0 \rightarrow \lambda\mu_0~, \qquad \mu \rightarrow \lambda^{-1} \mu~.
\fe
Alternatively, we can fix the coupling constants, and transform the scalar field as
\ie
\phi \rightarrow \lambda^{1/2} \phi~.
\fe
This means the radius of the scalar becomes larger as the theory flows to the IR, i.e., the scalar becomes non-compact. This is the same as in the 2+1d theory we reviewed in Section \ref{sec:2+1dXY}, and it is to be contrasted with the 2+1d $\phi$-theory \eqref{eq:action_continuum}, which is invariant under the same scale transformation.

From \eqref{phi-lat-energy}, we see that the momentum and winding states are infinitely heavy in this limit \cite{paper2}, which is the same as in the similar 2+1d theory.  Hence, these two symmetries are not spontaneously broken even if we later take the infinite volume limit $\ell\to \infty$. The fact that the symmetry is broken in one way of taking the infinite volume limit and it is unbroken in another way of taking this limit is again a sign of UV/IR mixing.

The infinite energy of charged states has another consequence.  The local operators charged under the momentum symmetry do not act in the low-energy theory. Therefore, the continuum $\phi$-theory is robust under perturbations by such local operators. Operators charged under the winding symmetry are not local, so the winding symmetry is also robust. See Table \ref{tbl:cont-SSB-robustness} for a comparison with other theories.

Another way to see that the momentum subsystem  symmetry is unbroken is to consider the monopole two-point function\footnote{The spatial positions of both monopoles must be the same because of the subsystem symmetry.},
\ie
&\langle e^{i\phi(0,0,\tau)} e^{-i\phi(0,0,0)} \rangle \sim \exp\left[ -\frac{|\tau|}{2\mu_0 \ell_x \ell_y \ell_z} \left( \frac{\ell_x}{a} + \frac{\ell_y}{a} + \frac{\ell_z}{a} - 2 \right) \right]~,
\fe
The leading contribution is from the exchange of the lowest energy state created by $e^{i\phi}$. The two-point function vanishes in the $a \rightarrow 0$ limit, and hence, the monopole operator does not act in the continuum. In particular, the $U(1)$ momentum subsystem symmetry, under which $e^{i\phi}$ is charged, is not spontaneously broken.

\begin{table}[t]
\begin{center}
\begin{tabular}{|c|c|c|c|}
\hline
Theory & Light states & SSB & Robustness
\tabularnewline
\hline
3+1d $\phi$-theory & Plane waves & No & Yes
\tabularnewline
3+1d $\hat A$-theory & Plane waves & No & Yes
\tabularnewline
\hline
3+1d $\hat \phi$-theory & Plane waves and winding & No & Yes
\tabularnewline
3+1d $A$-theory & Plane waves and electric & No & Yes
\tabularnewline
\hline
\end{tabular}
\caption{The spectrum of the theories of Sections \ref{sec:3+1dXYplaq} and \ref{sec:3+1dtensor} in infinite volume with fixed continuum coupling constants. As in Table \ref{tbl:latticelargeL2}, we note whether the momentum/magnetic symmetry is spontaneously broken in the infinite volume limit and whether the continuum theory is robust under deformations by winding/magnetic symmetry violating operators.}\label{tbl:cont-SSB-robustness}
\end{center}
\end{table}

\subsection{3+1d $\hat \phi$-theory/$A$-theory}\label{sec:3+1dtensor}

The modified Villain action of the 3+1d $\hat \phi$-theory is \cite{Gorantla:2021svj}
\ie\label{3+1d-hatphi-modVill-action}
&S= \frac{\hat \beta_0}{12}\sum_{\text{dual }\tau\text{-link}}~\sum_{i\ne j\ne k} (\Delta_\tau \hat \phi^{k(ij)} - 2\pi \hat n^{k(ij)}_\tau)^2 + \frac{\hat \beta}{4} \sum_{i\ne j\ne k}~\sum_{\text{dual }k\text{-link}} (\Delta_k \hat \phi^{k(ij)} - 2\pi \hat n^{ij})^2
\\
& \quad + \frac i2 \sum_{i\ne j\ne k}~\sum_{ij\text{-plaq}} A_{ij} (\Delta_\tau \hat n^{ij} - \Delta_k \hat n_\tau^{k(ij)}) + i \sum_{\tau\text{-link}} A_\tau \sum_{i<j} \Delta_i \Delta_j \hat n^{ij}~,
\fe
where the fields obey the constraints
\ie
&\hat \phi^{x(yz)}+\hat \phi^{y(zx)}+\hat \phi^{z(xy)}=0~,
\\
&\hat n_\tau^{x(yz)}+\hat n_\tau^{y(zx)}+\hat n_\tau^{z(xy)}=0~.
\fe
Here $\hat \phi^{k(ij)}$ are real fields, $(\hat n^{k(ij)}_\tau,\hat n^{ij})$ are integer-valued gauge fields, and $(A_\tau,A_{ij})$ are Lagrange multipliers imposing the flatness of the gauge field $(\hat n^{k(ij)}_\tau,\hat n^{ij})$.  A duality transformation makes $ A$ the dynamical field and turns $\hat \phi$ into a Lagrange multiplier \cite{Gorantla:2021svj}.  We refer to the corresponding dual continuum theory as the $ A$-theory.

The continuum field theory of this lattice model is described by the action\cite{paper2}
\ie\label{3dphihattheory}
\int d\tau dx dy dz ~ \sum_{i\ne j\ne k} \left[ \frac{\hat \mu_0}{12} (\partial_\tau \hat\phi^{k(ij)})^2 + \frac{\hat \mu}{4} (\partial_k \hat \phi^{k(ij)})^2 \right]~,
\fe
where $\hat \phi^{k(ij)}$ are compact scalars that obey the constraint $\hat \phi^{x(yz)}+\hat \phi^{y(zx)}+\hat \phi^{z(xy)}=0$. This corresponds to setting $a_\tau=a$, and the lattice parameters as
\ie
\hat \beta_0 = \frac{\hat \mu_0 a^3}{a_\tau}= {\hat \mu_0 a^2}~, \qquad \hat \beta = \hat \mu a_\tau a =\hat \mu a^2~,
\fe
where $\hat \mu_0,\hat \mu$ are fixed continuum coupling constants both with mass dimension +2.
The continuum $\hat\phi$-theory is dual to the $A$-theory of \cite{paper2}.
See also \cite{Xu2008,Slagle:2017wrc,Bulmash:2018lid,Ma:2018nhd,You:2018zhj,Radicevic:2019vyb} for related discussions on this tensor gauge theory of $(A_\tau,A_{ij})$.

Following an analysis similar to that in Section \ref{sec:spectrum},  we find the spectrum of this lattice model:
\ie\label{hatphi-lat-energy}
&{\rm plane\ waves}\qquad \qquad &&E_\text{wave} \sim \frac{1}{a_\tau L}\sqrt{\frac{\hat \beta}{\hat \beta_0}}=\frac{1}{\ell}\sqrt{\frac{\hat \mu}{\hat \mu_0}} ~,\\
&{\rm momentum\ states}\qquad &&E_\text{mom} \sim \frac{1}{a_\tau L^2\hat \beta_0}=\frac{1}{a \ell^2 \hat \mu_0}~,\\
&{\rm winding\ states} \qquad \quad &&E_\text{wind} \sim \frac{\hat \beta}{a_\tau}=a \hat \mu~.
\fe
See Table \ref{tbl:lattice-energy} and \ref{tbl:cont2-energy} for a comparison with other theories.

Similar to the discussion in Section \ref{sec:2+1dXY}, we will consider two different low-energy limits: one with the lattice couplings fixed, and another with the continuum couplings fixed. We will discuss spontaneous symmetry breaking and robustness of these low-energy theories.

\subsubsection{Fixed lattice couplings}\label{filachatphi}

In the fixed lattice coupling limit, we take $L\rightarrow \infty$ while keeping the lattice parameters fixed. It follows from \eqref{hatphi-lat-energy} that in this limit, the momentum states are the lightest states. (We will soon discuss the fixed continuum coupling limit and then the winding states will be the lightest states.) Therefore, the momentum symmetry is spontaneously broken in the infinite volume theory.

Perturbing the theory by local operators charged under the momentum symmetry changes the low-energy spectrum, so the momentum symmetry is not robust.\footnote{This agrees with the conclusion in \cite{Xu2008}, which analyzed a closely related tensor gauge theory and stressed the lack of robustness of its magnetic symmetry.} Operators charged under the winding symmetry are not point-like, so the winding symmetry is robust.   See Table \ref{tbl:latticelargeL2} for a comparison with other theories.

\subsubsection{Fixed continuum couplings}

In the fixed continuum coupling limit \cite{paper2}, we take the limit $a\rightarrow 0$ while keeping the physical size of the system $\ell=aL$ and the continuum coupling constants $\hat\mu_0,\hat\mu$ fixed. In this limit, the theory is described by the continuum $\hat\phi$-theory \eqref{3dphihattheory}.

The continuum theory is not scale-invariant. Under the scale transformation $x^i \rightarrow \lambda x^i$ and $\tau \rightarrow \lambda \tau$, the coupling constants transform as
\ie
\hat \mu_0 \rightarrow \lambda^2\hat \mu_0~, \qquad \hat \mu \rightarrow \lambda^2 \hat \mu~.
\fe
Alternatively, we can fix the coupling constants, and transform the scalar fields as
\ie
\hat \phi^{k(ij)} \rightarrow \lambda \hat \phi^{k(ij)}~.
\fe
This means the radius of the scalars becomes larger as the theory flows to the IR, i.e., the scalars becomes non-compact. This is the same as in the 2+1d theory we reviewed in Section \ref{sec:2+1dXY}.

From \eqref{hatphi-lat-energy}, we see that the momentum states are infinitely heavy in this limit. Hence the momentum symmetry is not spontaneously broken. Relatedly, the continuum $\hat \phi$-theory, or equivalently the continuum $A$-theory of \cite{paper2}, is robust under perturbations by local operators charged under the momentum/magnetic symmetry.\footnote{Note that this conclusion is the opposite of that in the limit with fixed lattice coupling of Section \ref{filachatphi}, again reflecting the subtlety in the continuum limit and the UV/IR mixing.}
Moreover, operators charged under the winding/electric symmetry are not point-like local operators, so the winding/electric symmetry is also robust. See Table \ref{tbl:cont-SSB-robustness} for a comparison with other theories.

\section{Conclusions}\label{conclusions}

Most of this paper discussed the 2+1d XY-plaquette model, whose lattice Hamiltonian is \cite{PhysRevB.66.054526}
\ie\label{eq:iH_XY_plaqC}
H  = { \mathcal{U} \over2} \sum_{\text{site}} \pi^2 - {\mathcal{K} }\sum_{xy-\text{plaq}}\cos(\Delta_x\Delta_y\phi)~.
\fe
We were particularly interested in its gapless phase, which seems to be described by the Euclidean continuum theory based on the action \cite{PhysRevB.66.054526}
\ie\label{eq:i action_continuumC}
S=\int d\tau dx dy\left[\frac{\mu_0}{2}(\partial_\tau\phi)^2+\frac{1}{2\mu}(\partial_x\partial_y\phi)^2\right]~.
\fe

We had two related goals.  First, since the analysis of the continuum model \eqref{eq:i action_continuumC} in \cite{paper1} was quite subtle, we wanted to make it more precise.  Second, we wanted to clarify the relation between this continuum model and the lattice model \eqref{eq:iH_XY_plaqC}.

As a first step, we replaced the lattice system \eqref{eq:iH_XY_plaqC} with the modified Villain lattice theory based on the Euclidean action
 \cite{Gorantla:2021svj}
\ie\label{eq:iS-Villain-latticeC}
S=\frac{\beta_0}{2}\sum_{\tau\text{-link}}(\Delta_\tau\phi-2\pi n_\tau)^2+\frac{\beta}{2}\sum_{xy\text{-plaq}}(\Delta_x\Delta_y\phi-2\pi n_{xy})^2+i\sum_{\text{cube}}\phi^{xy}(\Delta_\tau n_{xy}-\Delta_x\Delta_y n_\tau)~.
\fe
At weak coupling, i.e., $\mathcal{K}\gg \mathcal{U}$ or equivalently $\beta_0,\beta\gg 1$ these two lattice models are essentially the same.  (The relations between their parameters were spelled out in the introduction.)  However, the lattice model \eqref{eq:iS-Villain-latticeC} has all the internal symmetries of the continuum model \eqref{eq:i action_continuumC} and it is free. Placing this model on a finite lattice (with $L$ sites along each spatial direction), we have a completely rigorous setup. It gives us a regularized version of \eqref{eq:i action_continuumC}. Furthermore, since it is free, it can be analyzed exactly.  Then, we explored various limits of this lattice model and clarified the relation between it and the continuum theory \eqref{eq:i action_continuumC}.

At weak coupling, the spectrum of all these theories includes various kinds of excitations with energies of the following order of magnitude
\ie\label{latticespectrumC}
&{\rm plane\ waves}\qquad \qquad &&{\sqrt{\mathcal UK}\over L^2} \approx{1\over \sqrt{\mu_0 \mu} a^2 L^2} ~,\\
&{\rm momentum\ states}\qquad &&{{\mathcal U}\over L}\approx{1\over \mu_0 a^2L}~,\\
&{\rm winding\ states} \qquad \quad &&{{\mathcal K}\over L}\approx{1\over \mu a^2L}~.
\fe

We considered two different limits of the lattice model, both of them seem to lead to the continuum action \eqref{eq:i action_continuumC} with finite $\mu$ and $\mu_0$.

The continuum limit involves
\ie
L\to \infty \quad , \quad a \to 0 \quad {\rm with\ fixed}\quad \ell=aL\quad, \quad \mu_0 \quad ,\quad \mu
\fe
and therefore,
\ie
{\mathcal U} \approx {1\over \mu_0 a^2} \to \infty \quad, \quad {\mathcal K}\approx {1\over \mu a^2}\to \infty ~.
\fe
The fact that ${\mathcal U}$ and ${\mathcal K}$ are taken to infinity, signifies that this is a low-energy limit.  In this limit, the momentum and winding states are pushed out of the spectrum and only the plane waves remain.
The resulting continuum field theory is described by the action \eqref{eq:i action_continuumC} with finite couplings $\mu,\mu_0$. It exists in finite volume (finite $\ell$), which can later be taken to infinity.

Another interesting limit is the thermodynamic limit:
\ie
L\to \infty \quad {\rm with\ fixed}\quad a \quad , \quad {\mathcal U} \quad , \quad  {\mathcal K}
\fe
and therefore,
\ie
\ell \to \infty~.
\fe
In this limit, the energies of all the states in \eqref{latticespectrumC} go to zero.

These two limits can be considered in any lattice system.  We can first take the continuum limit to find a finite volume continuum field theory and then take its infinite volume limit.  We can also take the thermodynamic limit and then take the low-energy limit, i.e., scale $a\to 0$.  Usually, these two limits commute.  This is demonstrated in the review of the ordinary 1+1d XY model in Appendix \ref{onepone}.

The novelty here is that these two limits do not commute.  This is clear already from the form of the spectrum \eqref{latticespectrumC} and was discussed in detail in our analysis of the correlation functions.

Which order of limits is the right one?  This depends on the problem we are interested in.  In a given lattice model $L$ is finite and $a$ is nonzero.  Then, the various limits are approximations of the finite problem.  Depending on the energies we are interested in, one of these limits might be a more appropriate approximation than the other.

The dichotomy between these two limits was present already in our discussion in \cite{paper1}.  There, the continuum limit, which focuses only on the plane waves was referred to as the conservative approach.  In \cite{paper1}, we also entertained the possibility of keeping also the charged states and referred to this option as the ambitious approach.  We now identify the latter as the thermodynamic limit.

As we saw in the discussion of the spectrum and the correlation functions in the body of the paper, the continuum limit leads to a peculiar, but almost standard, quantum field theory.  However, the thermodynamic limit leads to quite a singular theory.  First, $\ell$ is infinite and we cannot consider the theory in finite volume.  Second, as explained above, it is not scale invariant -- the momentum and winding states do not respect the scaling symmetry \eqref{scalingxy}.  More generally, the correlation functions of this theory are quite puzzling, as they cannot be renormalized in the usual way.  Also, as stressed in \cite{paper1}, some observables computed in this effective theory are not universal -- certain high derivative terms can change the results associated with the charged states.

\section*{Acknowledgements}

We thank F.\ Burnell, S.\ Sachdev, and C.\ Xu for helpful discussions. PG was supported by Physics Department of Princeton University.  HTL was supported
by a Centennial Fellowship and a
Charlotte Elizabeth Procter Fellowship from Princeton University and Physics Department of Princeton University. The work of NS was supported in part by DOE grant DE$-$SC0009988.  NS and SHS were also supported by the Simons Collaboration on Ultra-Quantum Matter, which is a grant from the Simons Foundation (651440, NS).
Opinions and conclusions expressed here are those of the authors and do not necessarily reflect the views of funding agencies.

\appendix

\section{Correlation functions in the 1+1d XY model}\label{onepone}

This appendix reviews well known computations of correlation functions in the standard 1+1d XY model.  We include it here in order to demonstrate our notation and to contrast these expressions with those of the XY-plaquette model in Section \ref{sec:correlation}.

We work in the Hamiltonian formalism, where the time is continuous and the space is a discrete one-dimensional lattice with $L$ sites with periodic boundary conditions. The Hamiltonian is
\ie\label{XY1p1}
H  = { \mathcal U \over2} \sum_{\hat x=1}^L \pi_{\hat x}^2 - \mathcal K \sum_{\hat x=1}^L \cos (\Delta_x \phi_{\hat x})\,,
\fe
where $\pi_{\hat x}$ and $\phi_{\hat x}$ obey $[\phi_{\hat x} , \pi_{\hat x'} ] = i \delta_{\hat x, \hat x'}$ and $\Delta_x \phi_{\hat x} = \phi_{\hat x+1} -\phi_{\hat x}$.
Here $\phi_{\hat x}$ is circle-valued  and $\pi_{\hat x}$ is integer-valued.
The mass dimensions of $\mathcal U,\mathcal K$ are both +1.

The phase diagram of this model is controlled by the dimensionless parameter $\mathcal K/\mathcal U$.
At large $\mathcal K/\mathcal U$ the model is gapless and it is described by the continuum compact boson CFT, while at small $\mathcal K/\mathcal U$ the model is gapped.
We will assume large $\mathcal K/\mathcal U$ so that we are in the gapless phase.
In this limit, we can approximate the cosine by a quadratic term and the Hamiltonian becomes
\ie\label{eq:XYfree}
H  = {  U \over2} \sum_{\hat x=1}^L \pi_{\hat x}^2 +  \frac{K}{2} \sum_{\hat x=1}^L (\Delta_x \phi_{\hat x})^2\,,
\fe
with $U\approx\mathcal{U}$ and $K\approx \mathcal{K}$.  Here we omitted the integer-valued gauge fields that are needed to preserve the compactness of the field $\phi_{\hat x}$. In fact, the more precise version of the Hamiltonian \eqref{eq:XYfree} is essentially the Hamiltonian of the modified Villain version of the Hamiltonian \eqref{XY1p1}  \cite{Gorantla:2021svj}.  Therefore, from this point on we will consider  the Hamiltonian \eqref{eq:XYfree} for all values of $K/U$.

The model \eqref{eq:XYfree} is exactly solvable.
It has plane wave states with dispersion relation
\ie
\omega_n =  2\sqrt{ UK}  \left|\sin \left({\pi n\over L}\right)\right|~
\fe
and momentum and winding states with energies
\ie
E_{\text{mom}}\sim\frac{U}{L}~,\quad
E_{\text{wind}}\sim\frac{K}{L}~.
\fe
These values are also present in Table \ref{tbl:lattice-energythr}.

Let us compute the two-point function
\ie
\langle e^{i\phi_{\hat x} (\tau) } e^{-i \phi_{0} (0)}\rangle= e^{ - F(\hat x,\tau)~.}
\fe
The propagator of the model \eqref{eq:XYfree} is
\ie
F(\hat x,\tau) & = - \langle \phi_{\hat x}(\tau) \phi_0(0) \rangle
\\
&= { U \over2 L}|\tau | - {1\over 4L }  \sqrt{U\over K}\sum_{-{L\over2} <n \le {L\over2}\atop n\neq 0} {1\over |\sin({\pi n\over L})|}  e^{-2\sqrt{UK} |\sin({\pi n\over L})| |\tau|} \cos\left(\frac{2\pi n \hat x}{L}\right)
~.
\fe

In addition to $K/U$ and $L$, there are two other dimensionless parameters: $\hat x$, and  $T\equiv 2\sqrt{UK} |\tau| $.
As in the discussion in  Section \ref{sec:correlation}, we will be interested in the following limits:
\begin{itemize}
\item Continuum limit:
\ie
&1 \ll T ,\hat x \sim L\,,\qquad U/K=\text{fixed}\,.
\fe
 To understand this limit better, we write $U,K,L, T,\hat x$ in terms of the continuum variables $R,\ell,\tau,x$:
\ie
U = {2\pi \over R^2 a}\,,\qquad K = {R^2\over 2\pi a} \,,\qquad
L  ={ \ell\over a}\,,\qquad T=  {2|\tau|\over a}\,,\qquad \hat x = \frac{x}{a}\,,
\fe
where $R$ is the radius of the continuum compact boson CFT.
This   limit is  equivalent to taking $a\to0$, while holding the other continuum variables  fixed.  It leads to the continuum correlation function  in finite volume $\ell$.

\item Thermodynamic limit: \ie
&1\sim T,\hat x \ll L\,,
\fe
 This gives the correlation function in infinite space.
 \end{itemize}

\paragraph{Continuum limit}

In this limit, the propagator becomes
\ie
F(\hat x,\tau)
&\to { \pi \over \ell R^2}|\tau |- {1\over  R^2 }  \sum_{n=1}^{\infty} {1\over n}  e^{-2 \pi {n\over \ell} |\tau|} \cos\left( \frac{2\pi n x}{\ell} \right)
\\
&= { \pi \over \ell R^2}|\tau | +  {1\over 2R^2} \log \left[ 1- 2\cos\left(\frac{2\pi x}{\ell}\right)e^{-{2 \pi \over \ell} |\tau|} + e^{-{4 \pi \over \ell} |\tau|} \right]
\,.
\fe

For $|\tau|\gg \ell$, the first term in $F(\hat x, \tau)$ dominates and we have
\ie
\langle e^{i\phi_{\hat x} (\tau) } e^{-i \phi_{0} (0)}\rangle \sim e^{ - {\pi \over R^2\ell}|\tau| }~.
\fe
This describes the propagation of the lowest energy states created by the operator $e^{i\phi}$.

Alternatively, for $x,|\tau|\ll \ell$, i.e., in the infinite volume limit, we obtain the famous power-law two-point function:
\ie\label{powerlaw}
\langle e^{i\phi_{\hat x} (\tau) } e^{-i \phi_{0} (0)}\rangle \propto {1\over (\tau^2 + x^2)^{1\over 2R^2} } \,.
\fe

\paragraph{Thermodynamic limit}

In this limit, we can approximate the sum by an integral:
\ie
{1\over L}  \sum_n \to {1\over 2\pi} \int_{-\pi }^\pi dk \,.
\fe
Then the propagator becomes
\ie
F(\hat x,\tau) \to {1\over 2\pi} {1\over 4} \sqrt{ U\over K}
\int_{-\pi }^\pi dk~{1\over |\sin ({k\over2})| }
\left(
1- e^{- |\sin( {k\over2} )|T}\cos(k\hat x)
\right)~,
\fe
where we have introduced a $\tau$- and $\hat x$-independent constant to regularize the integral. It can be absorbed into the wavefunction renormalization of the operator $e^{i\phi_{\hat x}}$.
The integral can be expressed in terms of the harmonic number $H_n$, and the regularized generalized hypergeometric functions:
\ie
&\int_{-\pi }^\pi dk~{1\over |\sin ({k\over2})| }
\left(
1- e^{- |\sin( {k\over2} )|T}\cos(k\hat x)
\right)
\\
& = 8 H_{2\hat x} - 4 H_{\hat x} + \frac{\pi}{2} \cos(\pi \hat x) T \left[ 2\sqrt{\pi}\,\,_2\tilde F_3 \left( \frac12,1; \frac32, 1-\hat x, 1+\hat x; \frac{T^2}{4} \right) \right.
\\
& \qquad \qquad \qquad \qquad \qquad \qquad \qquad \left. - T \,\,_2\tilde F_3\left( 1,1; 2, \frac32 - \hat x, \frac32 + \hat x; \frac{T^2}{4} \right) \right]~.
\fe

Next, we take $\hat x, T\gg 1$ with fixed $\hat x/T$ and obtain
\ie
F(\hat x,\tau) \to {1\over 4\pi} \sqrt{ U\over K}   \log\left(\frac{T^2}{4} + \hat x^2\right) \sim
{1\over 2R^2} \log \left( {\tau^2 + x^2 \over a^2} \right)~,
\fe
which reproduces the power-law two-point function \eqref{powerlaw} of the infinite-volume continuum theory.

We conclude that in the region $1\ll T , \hat x\ll L$, the lattice correlation function  gives the power-law decaying continuum correlation function \eqref{powerlaw} in the infinite volume limit. We can zoom into this region in two different ways: we first take the continuum limit $1\ll T , \hat x\sim L$ and then take the large volume limit $T/L, \hat x/L \ll 1$; or we first take the thermodynamic limit $1\sim T , \hat x\ll L$ and then the large separation limit $1\ll T , \hat x$.

This conclusion should be contrasted with the correlation functions in the 2+1d XY-plaquette model in Section \ref{sec:correlation}, where the continuum limit \eqref{contlimit} and the thermodynamic limit \eqref{thermolimit} do not have any overlapping region.

\section{Details of $\langle \partial_\tau\phi_{\hat x,\hat y}(\tau) \partial_\tau \phi_{0,0}(0)\rangle$ in the 2+1d XY-plaquette model}\label{sec:app_phidot2pt}

In this appendix, we compute the two-point function of $\partial_\tau \phi$ in both the continuum limit \eqref{contlimit} and the thermodynamic limit \eqref{thermolimit}.
The physical interpretation of these correlation functions was discussed in details in Section \ref{sec:phidot}.

\subsection{Continuum limit}\label{sec:app_continuum2pt}

In the continuum limit \eqref{contlimit}, the two-point function becomes \eqref{phidot-2pt-A.5}
\ie
\langle \partial_\tau\phi_{\hat x,\hat y}(\tau) \partial_\tau \phi_{0,0}(0)\rangle
\to -{8\pi^2 \over \mu^{1\over2}\mu_0^{3\over2}\ell_x^2 \ell_y^2 }  \sum_{n_x=1}^{\infty}\sum_{n_y=1}^{\infty} n_x n_y e^{-{4\pi^2\over \sqrt{\mu\mu_0}} {|\tau|\over \ell_x\ell_y} n_xn_y  }\cos\left(2\pi  n_x {x\over \ell_x}\right)\cos\left(2\pi  n_y{ y\over \ell_y}\right)~.
\fe

If we subsequently take the limit ${|\tau|\over \sqrt{\mu\mu_0}\ell_x\ell_y} \gg1$,  the sum is dominated by the four lightest plane wave states, which are represented by $(n_x,n_y)=(1,1)$:
\ie
&\langle \partial_\tau \phi_{\hat x,\hat y}(\tau) \partial_\tau \phi_{0,0}(0)\rangle
\approx
-{8\pi^2 \over\mu^{1\over2}\mu_0^{3\over2} \ell_x^2\ell_y ^2}
e^{-{4\pi^2\over \sqrt{\mu\mu_0}} {|\tau|\over \ell_x\ell_y} }\cos\left({2\pi  { x\over\ell_x}}\right)\cos\left({2\pi   {y\over\ell_y}}\right)~.
\fe

Alternatively, we can consider the infinite volume limit $\ell_i\to\infty$, while keeping $\tau,x,y$ fixed.
In this limit, we can replace the sums by integrals:
\ie
\frac{1}{\ell^i} \sum_{n_i=1}^\infty \rightarrow \frac{1}{2\pi}\int_0^\infty dp_i~,\qquad \text{where} \qquad \frac{2\pi n_i}{\ell^i} \rightarrow p_i~.
\fe
Then, we have
\ie
\langle \partial_\tau\phi_{\hat x,\hat y}(\tau) \partial_\tau \phi_{0,0}(0)\rangle \rightarrow& -{1 \over 2\pi^2\mu\mu_0}  \sqrt{\mu\over \mu_0}
\int_0^\infty dp_x \int_0^\infty dp_y~p_x p_ye^{- p_x p_y|\tau|\over \sqrt{\mu\mu_0}}\cos(p_x x)\cos(p_y y)
\\
&=-{1 \over 2\pi^2}\sqrt{\mu\over \mu_0}  \frac{1}{\tau^2} f\left(\frac{|\tau|}{\sqrt{\mu\mu_0}xy}\right)~,
\fe
where $f(\xi)$ is the following integral which can be expressed in terms of the Meijer G-functions
	\ie
	f(\xi)&=\xi^2 \int_0^\infty dk_x \int_0^\infty dk_y~k_x k_y e^{- k_x k_y \xi}\cos(k_x)\cos(k_y)
	\\
	&=2\sqrt{\pi}\xi^2\left[G^{2 1}_{13}\left(\frac{1}{4\xi^2}\left|\begin{smallmatrix}0\\\\1,1,\frac32\end{smallmatrix}\right.\right)-
	G^{21}_{13}\left(\frac{1}{4\xi^2}\left|\begin{smallmatrix}1\\\\1,2,\frac32\end{smallmatrix}\right.\right)\right]~.
	\fe
	
Let us consider various limits of the correlation function. For $\xi\ll1 $, $f(\xi)\approx\xi^2$ and we have
\ie\label{eq:phidot_continuum_tau=0}
\langle \partial_\tau \phi_{\hat x,\hat y}(\tau) \partial_\tau \phi_{0,0}(0)\rangle\approx-{1 \over 2\pi^2}\sqrt{\mu\over \mu_0}  \frac{1}{\mu\mu_0 x^2y^2}~,~~~~\frac{|\tau|}{\sqrt{\mu\mu_0}}\ll xy
\fe
For $\xi \gg1$, $f(\xi)\approx\log(\xi)$ and we have
\ie\label{phidot-2pt-xy-ne-0}
\langle \partial_\tau \phi_{\hat x,\hat y}(\tau) \partial_\tau \phi_{0,0}(0)\rangle\approx-{1 \over 2\pi^2}\sqrt{\mu\over \mu_0} \frac{1}{\tau^2} \log\left(\frac{|\tau|}{\sqrt{\mu\mu_0 }xy}\right)~,
~~~~xy\ll\frac{|\tau|}{\sqrt{\mu\mu_0}}
~.
\fe
We see here something interesting.
This expression diverges at finite $\tau$ when $xy\to 0$, which is a manifestation of the UV/IR mixing.  This is not the standard short-distance singularity.

Let us look at the case $xy=0$ more carefully with a finite but large $\ell_x,\ell_y$. We can set $ x=0$, while keeping $y$ nonzero in \eqref{phidot-2pt-A.5} (swapping $x$ and $y$ gives similar results).
\ie\label{phidot-2pt-x=0}
\langle \partial_\tau\phi_{0,\hat y}(\tau) \partial_\tau \phi_{0,0}(0)\rangle \rightarrow& -{1 \over 2\pi^2 \mu\mu_0}  \sqrt{\mu\over \mu_0} \int_{\frac{2\pi}{\ell_x}}^{\infty}dp_x\int_{\frac{2\pi}{\ell_y}}^{\infty}dp_y~p_x p_y e^{-  {p_xp_y|\tau|\over \sqrt{\mu\mu_0}}}\cos(p_y y)
\\
& = -{1\over 2\pi^2}  \sqrt{\mu\over \mu_0} \frac{1}{\tau^2}\int_{\frac{2\pi}{\ell_y}}^{\infty}dp_y~\frac{1}{p_y} e^{- p_y  {2\pi |\tau| \over \sqrt{\mu\mu_0} \ell_x}}\cos(p_y y)
\\
&= \frac{1}{4\pi^2}\sqrt{\frac{\mu}{\mu_0}} \frac{1}{\tau^2} \log\left[\left(\frac{2\pi y}{\ell_y}\right)^2+\left(\frac{4\pi^2|\tau|}{\sqrt{\mu\mu_0}\ell_x\ell_y}\right)^2\right]~.
\fe
The singularity at $x=0$ in \eqref{phidot-2pt-xy-ne-0} is now regularized by the IR cutoff $\ell_x, \ell_y$. Although we assumed $x = 0$, this result is approximately correct even for $x \ll {2\pi |\tau| \over \sqrt{\mu\mu_0} \ell_y}$.

Note that we can take the limit $\ell_x\rightarrow \infty$ in \eqref{phidot-2pt-x=0} and still find a finite answer. In other words, for $y\gg {2\pi |\tau| \over \sqrt{\mu\mu_0} \ell_x}$ and $x \ll {2\pi |\tau |\over \sqrt{\mu\mu_0} \ell_y}$, the result is
\ie\label{eq:phidot_continuum_x=0}
\langle \partial_\tau\phi_{\hat x,\hat y}(\tau) \partial_\tau \phi_{0,0}(0)\rangle \approx \frac{1}{2\pi^2}\sqrt{\frac{\mu}{\mu_0}} \frac{1}{\tau^2} \log\left(\frac{2\pi y}{\ell_y}\right)~.
\fe
On the other hand, the case $x=y=0$ requires both the IR cutoffs $\ell_x,\ell_y$. More generally, when $y \ll {2\pi| \tau| \over \sqrt{\mu\mu_0} \ell_x}$ and $x \ll {2\pi |\tau |\over \sqrt{\mu\mu_0} \ell_y}$, the result is
\ie\label{eq:phidot_continuum_xy=0}
\langle \partial_\tau\phi_{\hat x,\hat y}(\tau) \partial_\tau \phi_{0,0}(0)\rangle \approx \frac{1}{2\pi^2}\sqrt{\frac{\mu}{\mu_0}} \frac{1}{\tau^2} \log\left(\frac{4\pi^2|\tau|}{\sqrt{\mu\mu_0}\ell_x\ell_y}\right)~.
\fe
This agrees with equation (A.8) of \cite{paper1}.

\subsection{Thermodynamic limit}\label{sec:app_thermo2pt}

In the thermodynamic limit \eqref{thermolimit}, the two point function becomes
\ie
&\langle \partial_\tau \phi_{\hat x,\hat y}(\tau) \partial_\tau \phi_{0,0}(0)\rangle
\\
& \rightarrow -{UK\over 2\pi^2}\sqrt{U\over K}  \int_{-\pi}^{\pi} dk_x  \int_{-\pi}^{\pi} dk_y~
\left| \sin\left({k_x\over 2}\right) \sin\left({k_y\over 2}\right)\right|
e^{ - | \sin({k_x\over 2})\sin({k_y\over 2}) | T}\cos(k_x\hat x)\cos(k_y\hat y)~.
\fe
The integral can be expressed in terms of the regularized generalized hypergeometric functions
	\ie
	&\int_{-\pi}^{\pi} dk_x  \int_{-\pi}^{\pi} dk_y~\left| \sin\left({k_x\over 2}\right) \sin\left({k_y\over 2}\right)\right| e^{ - | \sin({k_x\over 2})\sin({k_y\over 2}) | T}\cos(k_x\hat x)\cos(k_y\hat y)
	\\
	&= \frac{\pi^2}{4}\cos(\pi \hat x)\cos(\pi \hat y)
	\\
	&\quad \times\left[4\,\,_3\tilde F_4\left(1,1,\frac{3}{2};\frac{3}{2}-\hat x,\frac{3}{2}+\hat x,\frac{3}{2}-\hat y,\frac{3}{2}+\hat y;\frac{T^2}{4}\right) \right.
	\\
	& \qquad\qquad\qquad - 4T \,\,_3\tilde F_4\left(\frac{3}{2},2,2;2-\hat x,2+\hat x,2-\hat y,2+\hat y;\frac{T^2}{4}\right)
	\\
	&
	\qquad\qquad\qquad+ \left.3T^2\,\,_3\tilde F_4\left(2,2,\frac{5}{2};\frac{5}{2}-\hat x,\frac{5}{2}+\hat x,\frac{5}{2}-\hat y,\frac{5}{2}+\hat y;\frac{T^2}{4}\right) \right]~.
	\fe

In the $T\gg 1$ limit, we have
\ie\label{phidot-2pt-lim-T>>1}
&\langle \partial_\tau \phi_{\hat x,\hat y}(\tau) \partial_\tau \phi_{0,0}(0)\rangle
\\
& \quad\approx -{UK\over 2\pi^2}\sqrt{U\over K} \frac{8}{T^2} \left[ 2\log\left(\frac{T}{4}\right) - \psi\left(\frac12+\hat x\right) - \psi\left(\frac12-\hat x\right) - \psi\left(\frac12+\hat y\right) - \psi\left(\frac12-\hat y\right) \right]~,
\fe
where $\psi(z)$ is the digamma function. If we further take $T\gg\hat x\hat y \gg 1$, we find
\ie\label{phidot-2pt-thirdlim-T>>xy}
\langle \partial_\tau \phi_{\hat x,\hat y}(\tau) \partial_\tau \phi_{0,0}(0)\rangle \approx -{UK\over 2\pi^2}\sqrt{U\over K} \frac{16}{T^2} \log\left(\frac{T}{4\hat x\hat y}\right) = -{1\over 2\pi^2}\sqrt{U\over K} \frac{1}{\tau^2} \log\left(\frac{\sqrt{UK}\tau}{\hat x\hat y}\right) ~.
\fe
In terms of the continuum variables, this expression agrees with \eqref{phidot-2pt-xy-ne-0}.
The fact that the two computations can be continued to each other is because the plane wave spectrum is continuous at energies of order $1/L$. The expression \eqref{phidot-2pt-thirdlim-T>>xy} is finite for $\hat x\hat y\neq 0$, but it diverges as $\hat x \hat y \rightarrow 0$. This divergence is resolved in the full expression \eqref{phidot-2pt-lim-T>>1}. Specifically, when $\hat x=0$, in the limit $T\gg \hat y\gg1$, we have
\ie
\langle \partial_\tau \phi_{\hat x,\hat y}(\tau) \partial_\tau \phi_{0,0}(0)\rangle
\approx -{UK\over 2\pi^2}\sqrt{U\over K} \frac{16}{T^2}\log\left(\frac{T}{ \hat y}\right)
~,
\fe
In terms of the continuum variables, it is
\ie
\langle \partial_\tau \phi_{\hat x,0}(\tau) \partial_\tau \phi_{0,0}(0)\rangle\approx -{1\over 2\pi^2}\sqrt{\mu\over \mu_0} \frac{1}{\tau^2}\log \left(\frac{4|\tau|}{\sqrt{\mu\mu_0}a y}\right) ~.
\fe
In contrast to \eqref{eq:phidot_continuum_x=0}, the expression is regulated by a UV cutoff $a$.
When both $\hat x=\hat y=0$, we have
\ie
\langle \partial_\tau \phi_{0,0}(\tau) \partial_\tau \phi_{0,0}(0)\rangle
\approx -{UK\over 2\pi^2}\sqrt{U\over K} \frac{16}{T^2}\log T ~.
\fe
In terms of the continuum variables, it is
\ie
\langle \partial_\tau \phi_{0,0}(\tau) \partial_\tau \phi_{0,0}(0)\rangle
\approx -{1\over 2\pi^2}\sqrt{\mu\over \mu_0} \frac{1}{\tau^2}\log \left(\frac{4|\tau|}{\sqrt{\mu\mu_0}a^2}\right)~,
\fe
which is also regulated by a UV cutoff $a$ in contrast to \eqref{eq:phidot_continuum_xy=0}.

On the other hand, in the limit $1\gg T$, the correlation function becomes
\ie
\langle \partial_\tau \phi_{\hat x,\hat y}(\tau) \partial_\tau \phi_{0,0}(0)\rangle \approx -{1\over 2\pi^2}\sqrt{U\over K} \frac{UK}{(\hat x^2-1/4)(\hat y^2-1/4)} ~.
\fe
If we further take the limit $\hat x\hat y\gg1\gg T$, we have
\ie
\langle \partial_\tau \phi_{\hat x,\hat y}(\tau) \partial_\tau \phi_{0,0}(0)\rangle \approx -{1\over 2\pi^2}\sqrt{U\over K} \frac{UK}{\hat x^2\hat y^2} ~.
\fe
In terms of the continuum variables, the expression agrees with \eqref{eq:phidot_continuum_tau=0}.
We could also consider the limit where $\hat x = 0$ and $\hat y \gg 1 \gg T$. In this case, the two-point function is
\ie
\langle \partial_\tau \phi_{0,\hat y}(\tau) \partial_\tau \phi_{0,0}(0)\rangle \approx {1\over 2\pi^2}\sqrt{U\over K} \frac{4UK}{\hat y^2} ~.
\fe

\section{Anomalies of the modified Villain model}\label{sec:anomaly_lattice}

In this appendix, we compute the 't Hooft anomaly of the modified Villain lattice model \eqref{eq:S-Villain-lattice}.

The modified Villain lattice model has a $U(1)$ momentum and a $U(1)$ winding subsystem symmetry.
There is a mixed 't Hooft anomaly between these two $U(1)$ symmetries.
It can be seen by coupling the momentum and winding subsystem symmetries to appropriate classical background gauge fields.  In the spirit of the Villain formulation, we represent them as  $(A_\tau,A_{xy};N_{\tau xy})$ and $(\tilde A^{xy}_\tau,\tilde A;\tilde N_\tau)$, where
$A_\tau,A_{xy}, \tilde A^{xy}_\tau,\tilde A$ are $\mathbb{R}$-valued and $N_{\tau xy} ,\tilde N_\tau$ are $\mathbb{Z}$-valued. The action is \cite{Gorantla:2021svj}:
\ie\label{appano}
S&=\frac{\beta_0}{2} \sum_{\tau\text{-link}} (\Delta_\tau \phi - A_\tau - 2\pi n_\tau)^2 + \frac{\beta}{2} \sum_{xy\text{-plaq}} (\Delta_x \Delta_y \phi - A_{xy} - 2\pi n_{xy})^2 \\
&\quad\quad+ i \sum_\text{cube} \phi^{xy} (\Delta_\tau n_{xy} - \Delta_x \Delta_y n_\tau + N_{\tau xy})
\\
&\quad\quad- \frac{i}{2\pi}\sum_{xy\text{-plaq}} \tilde A^{xy}_\tau (\Delta_x \Delta_y \phi-2\pi n_{xy}) - \frac{i}{2\pi}\sum_{\tau\text{-link}} \tilde A (\Delta_\tau \phi-2\pi n_\tau)- i \sum_\text{site} \tilde N_\tau \phi   ~,
\fe
with the gauge symmetry
\ie\label{backg}
&\phi \sim \phi + \alpha + 2\pi k~, && \phi^{xy} \sim \phi^{xy} + \tilde \alpha^{xy} + 2\pi k^{xy}~,
\\
&A_\tau \sim A_\tau + \Delta_\tau \alpha + 2\pi K_\tau~, && \tilde A^{xy}_\tau \sim \tilde A^{xy}_\tau + \Delta_\tau \tilde \alpha^{xy} + 2\pi \tilde K^{xy}_\tau~,
\\
&A_{xy} \sim A_{xy} + \Delta_x \Delta_y \alpha + 2\pi K_{xy}~, && \tilde A \sim \tilde A + \Delta_x \Delta_y \tilde \alpha^{xy} + 2\pi \tilde K~,
\\
&n_\tau \sim n_\tau + \Delta_\tau k - K_\tau~, && \tilde N_\tau \sim \tilde N_\tau + \Delta_\tau \tilde K - \Delta_x \Delta_y \tilde K^{xy}_\tau~.
\\
&n_{xy} \sim n_{xy} + \Delta_x \Delta_y k - K_{xy}~,
\\
&N_{\tau xy} \sim N_{\tau xy} + \Delta_\tau K_{xy} - \Delta_x \Delta_y K_\tau~,\qquad
\fe
Here, $K_\tau, K_{xy}, \tilde K^{xy}_\tau,\tilde K$ are integers, and $\alpha,\tilde \alpha^{xy}$ are real.  They are the classical gauge parameters of the classical background gauge fields $(A_\tau,A_{xy};N_{\tau xy})$ and $(\tilde A^{xy}_\tau,\tilde A;\tilde N_\tau)$ . The variation of the action under the background gauge transformation is
\ie\label{XYplaq-anomaly-lattice}
S&\to S+\frac{i}{2\pi} \sum_\text{site} \alpha (\Delta_\tau \tilde A-\Delta_x\Delta_y\tilde A_\tau^{xy}-2\pi\tilde N_\tau)
\\
&\qquad -i\sum_{\tau\text{-link}}K_{\tau}(\tilde A+\Delta_x\Delta_y \tilde\alpha^{xy})
-i\sum_{xy\text{-plaq}}K_{xy}(\tilde A_\tau^{xy}+\Delta_\tau\tilde\alpha^{xy})
+i\sum_{\text{cube}}N_{\tau xy} \tilde\alpha^{xy}~,
\fe
In the continuum limit, this anomalous gauge transformation becomes
\ie
 \frac{i}{2\pi}\int d\tau dx dy\, \alpha (\d_\tau \tilde A - \d_x \d_y \tilde A^{xy}_\tau ) ~.
\fe

\bigskip\centerline{\it Dipole deformation}\bigskip

Next, we  analyze the anomaly of the modified Villain lattice model after turning on the momentum dipole deformation \eqref{dipoledef}, as considered in Section \ref{sec:dipole-deform}.
The momentum subsystem symmetry is now broken to the ordinary  momentum symmetry subgroup.
We therefore set the background momentum subsystem gauge fields to zero $A_\tau=A_{xy} =N_{\tau xy}=0$,  and  couple the action  to classical background gauge fields for the ordinary momentum $U(1)$ symmetry.  Following the spirit of the modified Villain formulation, the latter gauge fields are $(A_\mu, N_{[\mu\nu]})$ with $\mu,\nu = \tau, x, y$.
Here $A_\mu$ is real and $N_{[\mu\nu]}$ is an integer satisfying
\ie\label{Nconst}
\epsilon^{\mu\nu\rho}\Delta_\mu N_{[\nu\rho]}=0~.
\fe

As is always the case, there is an ambiguity in this coupling.  It is related to the freedom in performing improvement transformations of the conserved currents.  We resolve some of this ambiguity by imposing the ${\mathbb Z}_4$ rotation symmetry generated by $(x,y)\to (-y,x)$.   We also demand that the action be invariant under the gauge transformations of the dynamical fields.

The action is then\footnote{Despite appearance, this action is invariant under the ${\mathbb Z}_4$ rotation symmetry generated by $(x,y)\to (-y,x)$.   This symmetry acts on all the fields in an obvious way (including in particular $N_{[xy]}\to N_{[xy]}$), except $n_{xy}\to -n_{xy} -N_{[xy]}$.  In checking it, recall the constraint \eqref{Nconst}.}
\ie\label{modviwithem}
S=&\frac{\beta_0}{2} \sum_{\tau\text{-link}} (\Delta_\tau \phi - A_\tau - 2\pi n_\tau)^2 + \frac{\beta}{4} \sum_{xy\text{-plaq}} (\Delta_x \Delta_y \phi - \Delta_x A_y - 2\pi n_{xy})^2
\\
& + \frac{\beta}{4} \sum_{xy\text{-plaq}} (\Delta_x \Delta_y \phi - \Delta_y A_x - 2\pi n_{xy} -2\pi N_{[xy]})^2\\
& -\varepsilon \sum_{x\text{-link}}\cos(\Delta_x\phi -A_x ) -\varepsilon \sum_{y\text{-link}}\cos(\Delta_y\phi -A_y )
+ i\sum_\text{cube} \phi^{xy} (\Delta_\tau n_{xy} - \Delta_x \Delta_y n_\tau +\Delta_x N_{\tau y})
\\
& - \frac{i}{2\pi}\sum_{xy\text{-plaq}} \tilde A^{xy}_{\tau} (\Delta_x \Delta_y \phi  - 2\pi n_{xy}-\pi N_{[xy]})
- \frac{i}{2\pi}\sum_{\tau\text{-link}} \tilde A (\Delta_\tau \phi - 2\pi n_\tau) - i \sum_\text{site} \tilde N_\tau \phi ~.
\fe
 The gauge transformations are
\ie\label{app-anom-gaugesymm}
&\phi \sim \phi + \alpha + 2\pi k~,
\\
& \phi^{xy} \sim \phi^{xy}+\tilde \alpha^{xy} + 2\pi k^{xy}~,
\\
&(n_\tau,n_{xy}) \sim (n_\tau + \Delta_\tau k - K_\tau,n_{xy} + \Delta_x \Delta_y k - \Delta_x K_y)~,
\\
&(A_\mu;N_{[\mu\nu]}) \sim (A_\mu+ \Delta_\mu \alpha+2\pi K_\mu; N_{[\mu\nu]}+\Delta_\mu K_\nu-\Delta_\nu K_\mu)~,
\\
& (\tilde A,\tilde A_\tau^{xy}) \sim (\tilde A + \Delta_x \Delta_y \tilde \alpha^{xy} + 2\pi \tilde K, \tilde A^{xy}_{\tau} +\Delta_\tau \tilde \alpha^{xy} + 2\pi \tilde K^{xy}_{\tau})~,
\\
&\tilde N_\tau \sim \tilde N_\tau + \Delta_\tau \tilde K - \Delta_x \Delta_y \tilde K^{xy}_{\tau} ~.
\fe
And the change in the action is
\ie\label{momwan}
S\to S+&\frac{i}{2\pi}\sum_{\text{site}}\alpha(\Delta_\tau \tilde A-\Delta_x\Delta_y\tilde A^{xy}_\tau-2\pi \tilde N_\tau)
-i\sum_{\tau\text{-link}}K_\tau(\tilde A+\Delta_x\Delta_y\tilde \alpha^{xy})
\\
&-\frac{i}{2}\sum_{xy\text{-plaq}}(\Delta_yK_x+\Delta_x K_y)(\tilde A^{xy}_\tau+\tilde K^{xy}_\tau+\Delta_\tau\tilde \alpha^{xy})
\\
&+\frac{i}{2}\sum_{\text{cube}}(\Delta_yN_{\tau x}+\Delta_xN_{\tau y})\tilde\alpha^{xy}+i\pi\sum_{xy\text{-plaq}}N_{[xy]}\tilde K_\tau^{xy}~.
\fe

In the continuum limit, this anomalous gauge transformation becomes
\ie
 \frac{i}{2\pi}\int d\tau dx dy\, \alpha (\d_\tau \tilde A - \d_x \d_y \tilde A^{xy}_\tau ) ~.
\fe
It signals the 't Hooft  anomaly between the ordinary momentum symmetry and the winding subsystem symmetry.

\bibliographystyle{JHEP}
\bibliography{phidraft}

\end{document}